\documentclass[pdflatex,sn-mathphys-num]{sn-jnl}

\usepackage{graphicx}%
\usepackage{multirow}%
\usepackage{amsmath,amssymb,amsfonts}%
\usepackage{amsthm}%
\usepackage{mathrsfs}%
\usepackage[title]{appendix}%
\usepackage{xcolor}%
\usepackage{textcomp}%
\usepackage{manyfoot}%
\usepackage{booktabs}%
\usepackage{algorithm}%
\usepackage{algorithmicx}%
\usepackage{algpseudocode}%
\usepackage{listings}%
\usepackage{bm}

\theoremstyle{thmstyleone}%
\theoremstyle{thmstyletwo}%

\theoremstyle{thmstylethree}%

\begin{document}

\title[Article Title]{Extremely strong spin-orbit coupling effect in light-element altermagnetic materials}


\author[1,3]{\fnm{Shuai} \sur{Qu}}
\author[1,3]{\fnm{Zhen-Feng} \sur{Ouyang}}
\author[1,2,3]{\fnm{Ze-Feng} \sur{Gao}}
\author[2]{\fnm{Hao} \sur{Sun}}
\author[1,3]{\fnm{Kai} \sur{Liu}}
\author*[1,3]{\fnm{Peng-Jie} \sur{Guo}}
\email{guopengjie@ruc.edu.cn}
\author*[1,3]{\fnm{Zhong-Yi} \sur{Lu}}
\email{zlu@ruc.edu.cn}

\affil[1]{
\orgdiv{School of Physics and Beijing Key Laboratory of Opto-electronic Functional Materials $\&$ Micro-nano Devices}. 
\orgname{Renmin University of China}, \city{Beijing}, 
\postcode{100872}, \country{China}}

\affil[2]{
\orgdiv{Gaoling School of Artificial Intelligence}, 
\orgname{Renmin University of China}, 
\city{Beijing}, \country{China}}

\affil[3]{
\orgdiv{Key Laboratory of Quantum State Construction and Manipulation (Ministry of Education)}, 
\orgname{Renmin University of China}, \city{Beijing}, 
\postcode{100872}, \country{China}}


\abstract{
Spin-orbit coupling is a key to realize many novel physical effects in condensed matter physics. Altermagnetic materials possess the duality of real-space antiferromagnetism and reciprocal-space ferromagnetism. It has not been explored that achieving strong spin-orbit coupling effect in light-element altermagnetic materials. In this work, based on symmetry analysis, the first-principles electronic structure calculations plus Dynamical Mean Field Theory, we demonstrate that there is strong spin-orbit coupling effect in light-element altermagnetic materials $\rm NiF_3$ and $\rm FeCO_3$, and then propose a mechanism for realizing such effective spin-orbit coupling. This mechanism reveals the cooperative effect of crystal symmetry, electron occupation, electronegativity, electron correlation, and intrinsic spin-orbit coupling. Our work provides an approach for searching light-element altermagnetic materials with an effective strong spin-orbit coupling.
}

\maketitle

\section{Introduction}\label{sec1}

Spin-orbit coupling (SOC) is ubiquitous in realistic materials and crucial for many novel physical phenomena emerging in condensed matter physics, including topological physics \cite{r1,r2,r3}, anomalous Hall effect \cite{r4}, spin Hall effect \cite{r5,r6}, magnetocrystalline anisotropy \cite{r7} and so on. For instance, quantum anomalous Hall (QAH) insulators are characterized by non-zero Chern numbers \cite{r8}. The Chern number is derived from the integration of Berry curvature over the occupied state of the Brillouin zone (BZ). For collinear ferromagnetic and antiferromagnetic systems, the integral of Berry curvature over the occupied state in the Brillouin zone must be zero without SOC due to the spin symmetry $\rm \{C_2^\perp T||T\}$. Here the $\rm C_2^\perp$ and $\rm T$ represent the 180 degrees rotation perpendicular to the spin direction and time-reversal operation, respectively. Therefore, QAH effect can only be realized in collinear magnetic systems when SOC is included \cite{r9,r10}. On the other hand, strong SOC may open up a large nontrivial bandgap, which is very important to realize QAH effect at high temperatures.  In general, strong SOC exists in heavy-element compounds. Unfortunately, the chemical bonds of heavy element compounds are weaker than those of light-element compounds, which leads to more defects in heavy-element compounds. Thus, the stability to realize exotic functionalities in heavy element compounds is relatively weak. 

An interesting question is whether the strong SOC effect can be achieved in light-element compounds. In 2008, Liu et al. proposed the correlation-enhanced spin-orbit effect in paramagnetic $\rm Sr_2RhO_4$ \cite{liu-prl}. Very recently, Li et al. demonstrated that the SOC can be enhanced in light-element ferromagnetic materials, which derives from the cooperative effect of crystal symmetry, electron occupancy, electron correlation, and intrinsic SOC \cite{r11}. This provides a new direction for the design of light-element materials with strong effective SOC.

Very recently, altermagnetism is proposed as a new magnetic phase possessing the duality of real-space antiferromagnetism and reciprocal-space ferromagnetism, distinct from ferromagnetism and conventional collinear antiferromagnetism \cite{r12,r13}. Moreover, altermagnetic materials have a wide range of electronic properties, which cover metals, semi-metals, semiconductors, and insulators \cite{r13,r14}. Different from ferromagnetic materials with $\bm s$-wave spin polarization, altermagnetic materials have $\bm k$-dependent spin polarization, which results in many exotic physical effects \cite{r12,r13,r15,r16,r17,r18,r19,r20,r21,r22}. With SOC, similar to the case of ferromagnetic materials, the time-reversal symmetry-breaking macroscopic phenomena can be also realized in altermagnetic materials \cite{r10,r23,r24,r25}. Nevertheless, altermagnetism is proposed based on spin group theory \cite{r26,r27,r28,r29} and the predicted altermagnetic materials have many light-element compounds \cite{r13,r14,r27}. Therefore, it is very important to propose a mechanism to enhance SOC in light-element compounds with altermagnetism and predict the corresponding compounds with strong SOC effect.

In this work, based on symmetry analysis, the first-principles electronic structure calculations based on density functional theory (DFT) plus Dynamical Mean Field Theory (DMFT), we predict that the light-element compounds $\rm NiF_3$ and $\rm FeCO_3$ are $ i$-wave altermagnetic materials with extremely strong SOC effect. Then, we further propose a mechanism to enhance SOC effect in light-element compounds with altermagnetism, which reveals the cooperative effect of crystal symmetry, electron occupation, electronegativity, electron correlation, and intrinsic SOC. We also explain the weak SOC effect in altermagnetic materials $\rm VF_3$, $\rm CrF_3$, $\rm FeF_3$ and $\rm CoF_3$.  

\begin{figure*}[htbp]
\centering
\includegraphics[width=1.00\textwidth]{figure/f1.jpg}
\caption{
The crystal structure and six collinear magnetic structures of $\rm NiF_3$. (a) and (b) are side and top views of the crystal structure, respectively. The cyan arrow represents the direction of easy magnetization axis. (c)-(h) are six different collinear magnetic structures including one ferromagnetic and five different collinear antiferromagnetic structures. The bond angle of Ni-F-Ni for the nearest neighbor Ni ions is 140 degrees. The primitive cell of $\rm NiF_3$ is shown in (d). The red and blue arrows represent spin-up and spin-down magnetic moments, respectively.
}\label{fig:1}
\end{figure*}

\section{Results}\label{sec2}
\begin{figure}[htbp]
\centering
\includegraphics[width=0.50\textwidth]{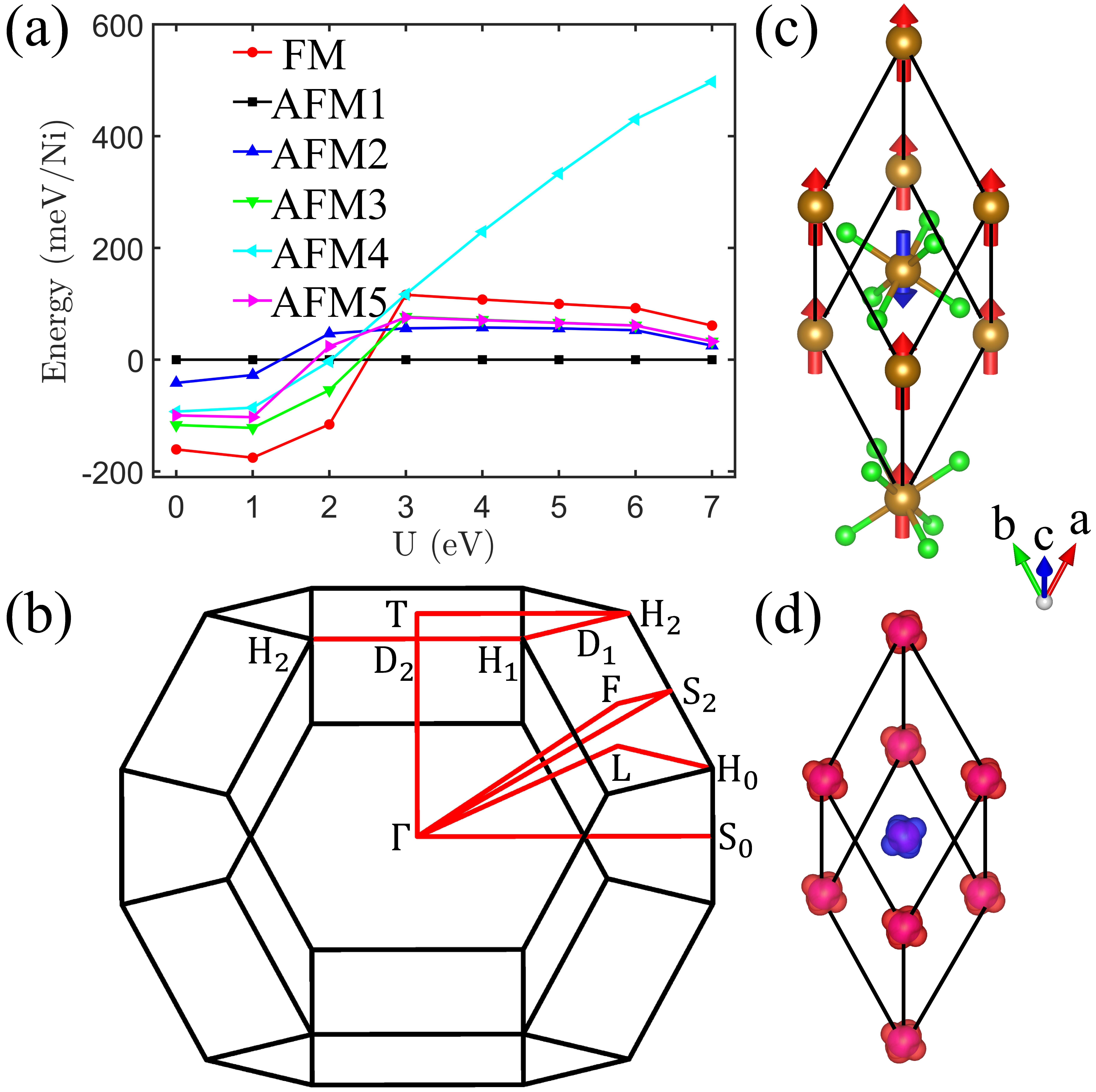}
\caption{
The magnetic ground state of $\rm NiF_3$ and the corresponding properties. (a) Relative energies of six different magnetic states with the variation of correlation interaction $\rm U$. (b) and (c) are the corresponding Brillouin zone and the magnetic primitive cell of $\rm NiF_3$, respectively. The high-symmetry lines and points are marked in the BZ. The red and blue arrows represent spin-up and spin-down magnetic moments, respectively. (d) The anisotropic polarization charge densities. The red and blue represent spin-up and spin-down polarization charge density, respectively.
}\label{fig:2}
\end{figure}

The $\rm NiF_3$ takes rhombohedral structure with nonsymmorphic $\rm R\overline{3}c$ (167) space group symmetry (Fig.\ref{fig:1}(a) and (b)). The corresponding elementary symmetry operations are $\rm C_{3z}$, $\rm C_2^1 t$ and $\rm I$, which yield the point group $\rm D_{3d}$. The $\rm t$ represents (1/2, 1/2, 1/2) fractional translation. To confirm the magnetic ground state of $\rm NiF_3$, we consider six different magnetic structures, including one ferromagnetic and five collinear antiferromagnetic structures which are shown in Fig.\ref{fig:1}(c)-(h). Then we calculate relative energies of six magnetic states with the variation of correlation interaction $\rm U$. With the increase of correlation interaction $\rm U$, the $\rm NiF_3$ changes from the ferromagnetic state to AFM1 state (Fig.\ref{fig:2}(a)). The AFM1 is of intralayer ferromagnetism and interlayer antiferromagnetism (Fig.\ref{fig:1}(d)), namely so-called G-type antiferromagnetic order in literature. In previous works, the correlation interaction $\rm U$ was selected as 6.7 $\rm eV$ for Ni 3$ d$ orbitals \cite{r30,r31}. Thus, the magnetic ground state of $\rm NiF_3$ is the AFM1 state, which is consistent with previous works \cite{r14}. On the other hand, since the bond angle of Ni-F-Ni for the nearest neighbor Ni ions is 140 degrees, the spins of the nearest neighbor and next nearest neighbor Ni ions are in antiparallel and parallel arrangement according to Goodenough-Kanamori rules \cite{r32}, respectively. This will result in $\rm NiF_3$ being the collinear antiferromagnetic state AFM1. Thus, the results of theoretical analysis are in agreement with those of theoretical calculation. 

Indeed, the AFM1 state is very simple and the corresponding magnetic primitive cell only contains two magnetic atoms. From Fig.\ref{fig:2}(c), the two Ni atoms with opposite spin arrangement are surrounded by F-atom octahedrons with different orientations, respectively. Thus, the two opposite spin Ni sublattices cannot be connected by a fractional translation. Due to two Ni ions located at space-inversion invariant points, the two opposite spin Ni sublattices cannot be either connected by space-inversion symmetry. However, the two opposite spin Ni sublattices can be connected by $\rm C_2^1 t$ symmetry. Thus, the $\rm NiF_3$ is an altermagnetic material. The BZ of altermagnetic $\rm NiF_3$ is shown in Fig.\ref{fig:2}(b) and both the high-symmetry lines and points are marked. In order to display the altermagnetic properties more intuitively, we calculate polarization charge density of altermagnetic $\rm NiF_3$. From Fig.\ref{fig:2}(d), the polarization charge densities of two Ni ions with opposite spin arrangement are anisotropic and their orientations are different, deriving from F-atom octahedrons with different orientations. The anisotropic polarization charge densities can result in $ k$-dependent spin polarization in reciprocal space. Moreover, according to different spin group symmetries, the $ k$-dependent spin polarization can form $ d$-wave, $ g$-wave, or $ i$-wave magnetism \cite{r12}. 

\begin{figure}[htbp]
\centering
\includegraphics[width=0.44\textwidth]{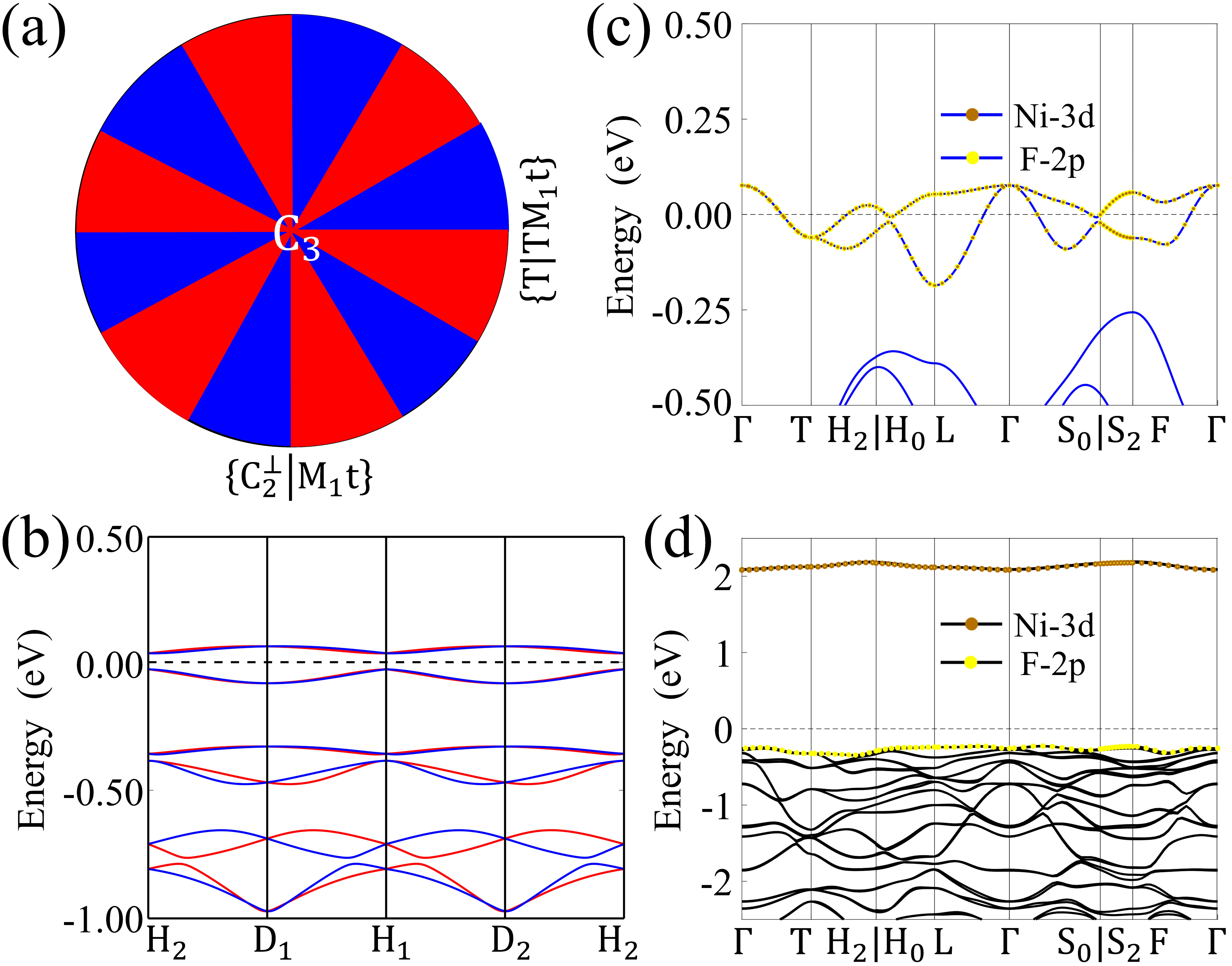}
\caption{
Schematic diagram of the $\bm i$-wave magnetism and electronic band structures of altermagnetic $\rm NiF_3$. (a) Schematic diagram of the $\bm i$-wave magnetism. The red and blue parts represent spin up and down, respectively. (b) The electronic band structure without SOC along the high-symmetry directions. The red and blue lines represent spin-up and spin-down bands, respectively. (c) and (d) are the electronic band structures without and with SOC along the high-symmetry directions.
}\label{fig:3}
\end{figure}

Without SOC, the nontrivial elementary spin symmetry operations in altermagnetic $\rm NiF_3$ have $\rm \{E||C_{3z} \}$, $\rm \{C_2^\perp ||M_1 t\}$, $\rm \{E||I\}$. The spin symmetries $\rm \{C_2^\perp ||M_1 t\}$, $\rm \{T||TM_1 t\}$, and $\rm \{E||C_{3z}\}$ make altermagnetic $\rm NiF_3$ being an $\bm i$-wave magnetic material, as shown in Fig. \ref{fig:3} (a). Meanwhile, the spins of bands are opposite along the $\rm H_2-D_1$ ($\rm H_2-D_2$) and $\rm D_1-H_2$ ($\rm D_2-H_1$) directions, reflecting features of $\bm i$-wave magnetism (Fig.\ref{fig:3}(b)).

In order to well understand the electronic properties, we also calculate the electronic band structures of altermagnetic $\rm NiF_3$. Without SOC, the $\rm NiF_3$ is an altermagnetic metal. There are four bands crossing the Fermi level due to spin degeneracy on the high-symmetry directions (Fig.\ref{fig:3} (c)). Especially, these four bands are degenerate on the $\rm \Gamma-T$ axis. In fact, any $ k$ point on the $\rm \Gamma-T$ axis has nontrivial elementary spin symmetry operations $\rm \{E||C_{3z}\}$ and $\rm \{C_2^\perp ||M_1 t\}$. And the spin symmetry $\rm \{E||C_{3z}\}$ has one one-dimensional irreducible real representation and two one-dimensional irreducible complex representations. Although the time-reversal symmetry is broken, altermagnetic materials can have equivalent time-reversal spin symmetry $\rm \{C_2^\perp T||IT\}$. The spin symmetry $\rm \{C_2^\perp T||IT\}$ will result in two one-dimensional irreducible complex representations to form a Kramers degeneracy. Meanwhile, the spin symmetry $\rm \{C_2^\perp ||M_1 t\}$ protects the spin degeneracy. Therefore, there is one four-dimensional and one two-dimensional irreducible representations on the $\rm \Gamma-T$ axis. The quadruple degenerate band crossing the Fermi level is thus protected by the spin group symmetry. Furthermore, the orbital weight analysis shows that these four bands are contributed by both the 3$\bm d$ orbitals of Ni and the $\bm p$ orbitals of F (Fig.\ref{fig:3} (c)). As is known to all, the F atom has the strongest electronegativity among all chemical elements, but the 2$ p$ orbitals of F do not fully acquire the 3$ d$-orbital electrons of Ni, which is very interesting. 

In our calculations, the number of valence electrons of $\rm NiF_3$ is 74, which makes the quadruple band only half-filled. This is the reason why the $\bm p$ orbitals of F do not fully acquire the $\bm d$-orbital electrons of Ni. When SOC is included, the spin group symmetry breaks down to magnetic group symmetry. The reduction of symmetry will result in the quadruple band to split into four non-degenerate bands. Since the F atom has the strongest electronegativity, the 2$\bm p$ orbitals of F will completely acquire the 3$\bm d$-orbital electrons of Ni. This will result in altermagnetic $\rm NiF_3$ to transform from metal phase to insulator phase. In order to prove our theoretical analysis, we calculate the electronic band structure of altermagnetic $\rm NiF_3$ with SOC. Just like our theoretical analysis, the 2$\bm p$ orbitals of F indeed fully acquire the 3$\bm d$-orbital electrons of Ni and altermagnetic $\rm NiF_3$ transforms into an insulator with a bandgap of 2.31 $\rm eV$ (Fig.\ref{fig:3} (d)). In general, the SOC strength of Ni is in the order of 10 $\rm meV$, so the effective SOC strength of altermagnetic $\rm NiF_3$ is two orders of magnitude higher than that of Ni. Thus, the SOC effect of altermagnetic $\rm NiF_3$ is extremely strong. 

\begin{figure}[htbp]
\centering
\includegraphics[width=0.44\textwidth]{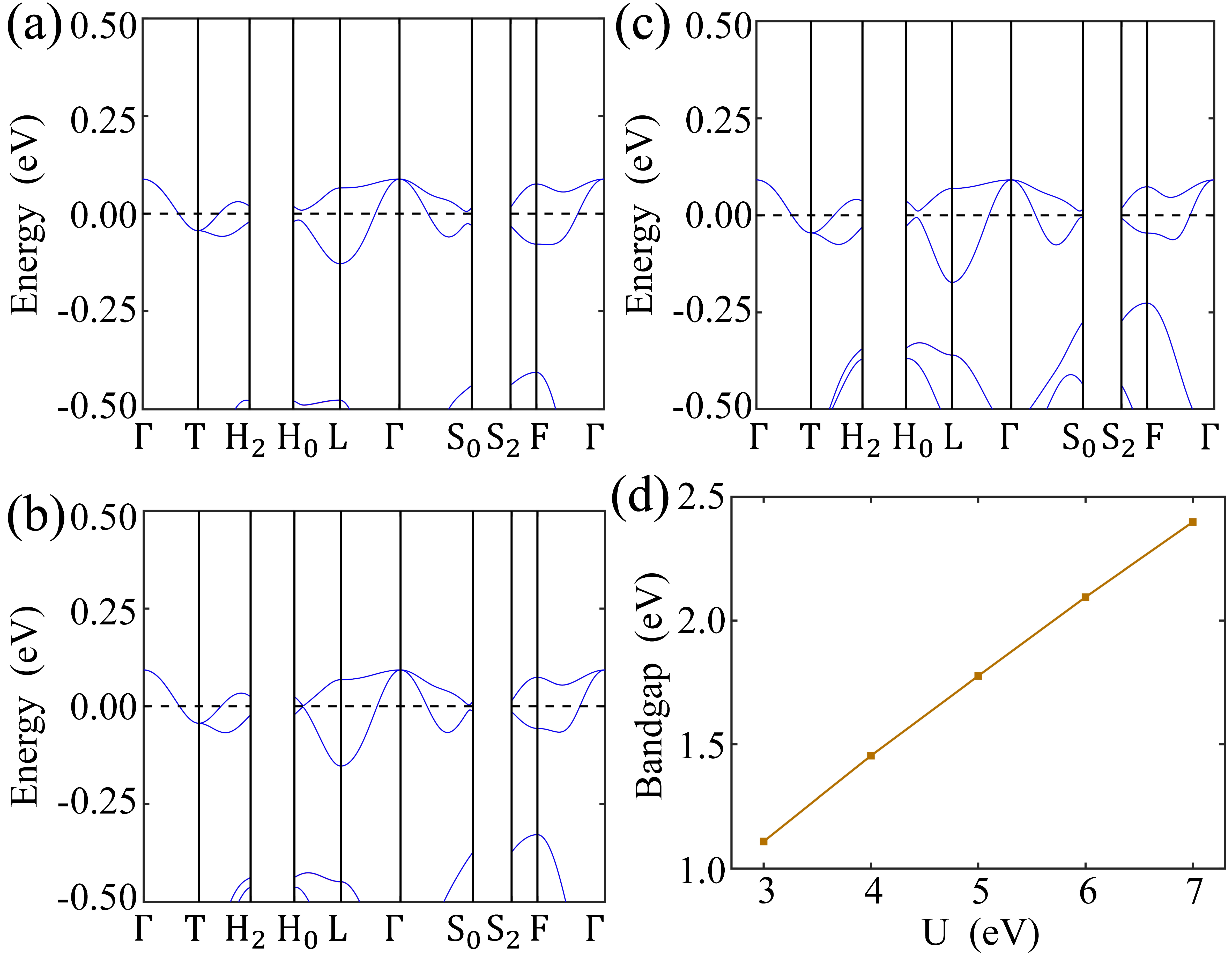}
\caption{
The electronic properties of altermagnetic $\rm NiF_3$ under different correlation interaction $\rm U$. (a), (b) and (c) are the electronic band structures along the high-symmetry directions without SOC under correlation interaction $\rm U = 3,5,7 eV$, respectively. (d) The bandgap as a function of correlation interaction $\rm U$ under SOC.
}\label{fig:4}
\end{figure}

In order to examine the effect of correlation interaction, we also calculate the electronic band structures of altermagnetic $\rm NiF_3$ under correlation interaction $\rm U=3,5,7 eV$, which are shown in Fig.\ref{fig:4} (a), (b), and (c), respectively. From Fig.\ref{fig:4} (a), (b) and (c), the correlation interaction has a slight effect on the band structure around the Fermi level without SOC, due to the constraints of spin symmetry and electron occupancy being 74. When including SOC, altermagnetic $\rm NiF_3$ transforms from a metal phase to an insulator phase under different correlation interaction $\rm U$. Moreover, the bandgap of altermagnetic $\rm NiF_3$ increases linearly with the correlation interaction $\rm U$ (Fig.\ref{fig:4} (d)). Thus, the correlation interaction can substantially enhance the bandgap opened by the SOC of altermagnetic materials. In order to make our results more reliable, we have also further completed the calculations respectively with hybrid functional and DFT+DMFT but without SOC. All these calculations are consistent with the calculated results of GGA+U, which are shown in Supplementary Material (SM). 

\section{Discussion}\label{sec12}
Now we well understand the reason for the extremely strong SOC effect in altermagnetic $\rm NiF_3$. A natural question is whether such a strong SOC effect can be realized in other altermagnetic materials. According to the above analysis, we propose four conditions for realizing such an effective strong SOC in light-element altermagnetic materials: First, the spin group of altermagnetic material has high-dimensional (greater than four dimensions) irreducible representation (crystal symmetry groups are presented in the SM \cite{r36}); Second, the band with high-dimensional representation crossing the Fermi level is half-filled by valence electrons; Third, non-metallic elements have strong electronegativity; Fourth, the altermagnetic material has strong electron correlation. To verify these four conditions, we further calculate the electronic band structures of four $ i$-wave altermagnetic materials ($\rm VF_3$, $\rm CrF_3$, $\rm FeF_3$, $\rm CoF_3$, and $\rm FeCO_3$), which have the same crystal structure and magnetic structure as $\rm NiF_3$ \cite{r14}. 
Our calculations show that these four materials ($\rm VF_3$, $\rm CrF_3$, $\rm FeF_3$, and $\rm CoF_3$) do not satisfy the second condition, while $\rm FeCO_3$ satisfies all four conditions. Therefore, the SOC effect of these four materials is weak (Fig. \ref{fig:6}), but the SOC effect of $\rm FeCO_3$ is extremely strong (Fig. \ref{fig:9}).  
On the other hand, since high-dimensional irreducible representations can be protected by spin space group in two-dimensional altermagnetic systems, the proposed mechanism is also applicable to two-dimensional light element altermagnetic materials, which may be advantage for realizing quantum anomalous Hall effect at high temperatures \cite{r10}. 

The mechanism for enhancing the SOC effect that we propose in altermagnetic materials is different from that in ferromagnetic materials \cite{r11}. First, the high-dimensional representation of the symmetry group is 2 or 3 dimensions in ferromagnetism, while in altermagnetism the high-dimension representation is 4 or 6 dimensions, so their symmetry requirements are entirely different. Second, the band with high-dimensional representation in ferromagnetism comes from $\bm d$ orbitals, while the band with high-dimensional representation in altermagnetism can come from the combination of $\bm p$ orbitals and $\bm d$ orbitals. Third, the enhancement of SOC effect derives from correlation interaction for ferromagnetic materials, but from both correlation interaction and the electronegativity of nonmetallic element for altermagnetic materials.  Due to one more degree of freedom to enhance the SOC effect, a stronger SOC effect can be achieved in the altermagnetic materials. Moreover, if the electronegativity of nonmetallic element is weak, different topological phases may be realized in altermagnetic materials when including SOC. On the other hand, the mechanism for enhancing SOC effect in altermagnetic materials can be also generalized to conventional antiferromagnetic materials. Due to the equivalent time-reversal symmetry, more spin groups with conventional antiferromagnetism have high-dimensional irreducible representations. Moreover, conventional antiferromagnetic materials are more abundant than altermagnetic materials, thus conventional light-element antiferromagnetic materials with strong SOC effect remain to be discovered.

In experiment, a recent experiment suggests that $\rm NiF_3$ is a semiconductor \cite{r33}, which is consistent with our calculations. This provides a strong evidence that $\rm NiF_3$ has a strong SOC effect. Different from $\rm NiF_3$, $\rm FeCO_3$ has not only been synthesized experimentally \cite{r48}, but also its magnetic structure has been confirmed by neutron scattering experiments \cite{r49}. Therefore, $\rm FeCO_3$ must have strong SOC effect, which can be confirmed by the APPES experiment and the first-principles electronic structure calculations. Finally, the metal phase of $\rm NiF_3$ is unstable, and both the Jahn-Teller and SOC effects can transform $\rm NiF_3$ into a stable insulator phase. Our calculations show that the SOC effect dominates the electronic properties of $\rm NiF_3$ rather than the Jahn-Teller effect (Detailed calculations and analysis are presented in the SM \cite{r36}).

\section{Conclusion}\label{sec13}

Based on symmetry analysis and the first-principles electronic structure calculations, we demonstrate that there is extremely strong SOC effect in altermagnetic materials $\rm NiF_3$ and $\rm FeCO_3$. Then, we propose a mechanism to enhance SOC effect in altermagnetic materials. This mechanism reveals the cooperative effect of crystal symmetry, electron occupation, electronegativity, electron correlation, and intrinsic spin-orbit coupling. The mechanism can explain not only the strong SOC effect in altermagnetic $\rm NiF_3$, but also the weak SOC in altermagnetic $\rm VF_3$, $\rm CrF_3$, $\rm FeF_3$, $\rm CoF_3$. Moreover, the mechanism for enhancing SOC effect can be also generalized to two-dimensional altermagnetic materials, being beneficial to realize  quantum anomalous Hall effect at high temperatures.

\section{Methods}\label{sec11}
The first-principles electronic structure calculations were performed within the framework of density functional theory (DFT) \cite{r37,r38} using the Vienna ab initio Simulation Package (VASP) \cite{r39,r40,r41} and Quantum ESPRESSO \cite{r52}. The generalized gradient approximation (GGA) of the Perdew-Burke-Ernzerhof (PBE) type \cite{r42} and the meta-GGA of the Strongly Constrained and Appropriately Normed (SCAN) type \cite{r51} were employed for the exchange-correlation functionals. The projector augmented wave (PAW) method \cite{r43,r44} was adopted to describe the interactions between valence electrons and nuclei. The kinetic-energy cutoff for the plane-wave basis was set to 600 $\rm eV$. Total energy convergence and atomic force tolerance were established at $10^{-6}$ $\rm eV$ and $10^{-4}$ $\rm eV/\AA$, respectively. To describe the Fermi-Dirac distribution function, a Gaussian smearing of 0.05 $\rm eV$ was used. The 16×16×16 and 8×8×4 Monkhorst-Pack $ k$ meshes were employed for BZ sampling of the unit cell and supercell, respectively. To account for the correlation effects of Ni 3$ d$ orbitals, we performed GGA+U \cite{r45} calculations by using the rotationally invariant approach introduced by Liechtenstein et al \cite{r50} and the simplified rotationally invariant version of Dudarev et al \cite{r46}. The empirical Hubbard $\rm U$ parameter for Ni 3$\bm d$ orbitals was determined to be 6.7 $\rm eV$ according to previous studies \cite{r30,r31,r47}.

The Heyd-Scuseria-Ernzerhof (HSE) functional \cite{r53} implemented in Quantum ESPRESSO  significantly enhanced the accuracy of electronic band structure calculations of altermagnetic $\rm NiF_3$ without SOC by integrating both local and non-local exchange-correlation contributions. The gygi-baldereschi approach was used for treating the Coulomb potential divergencies at small sampling vectors of the Fock operator.

Momentum-resolved spectral function obtained by the DFT+DMFT (dynamical mean-field theory) calculations without SOC for altermagnetic $\rm NiF_3$ at temperature $\rm T = 200 K$ by hybridization expansion continuous-time quantum impurity solver \cite{r54} based on the EDMFTF package \cite{r55}.And an “exact” double-counting scheme developed by Haule \cite{r56} was used. The real-frequency self-energy function was obtained by analytical continuation with the maximum entropy. Then it was used to calculate the momentum-resolved spectral function. And density-density form of the Coulomb repulsion was used. The 3$ d$ orbitals of Ni were considered to be correlated.

\backmatter

\section{Data availability}
Data are available from the corresponding authors (Peng-Jie Guo and Zhong-Yi Lu) upon reasonable request. We employed the density functional theory code VASP and EDMFTF, which can be obtained and purchased at http://www.vasp.at and http://hauleweb.rutgers.edu/, respectively.

\bibliography{sn-bibliography}


\begin{thebibliography}{56}
\ifx \bisbn   \undefined \def \bisbn  #1{ISBN #1}\fi
\ifx \binits  \undefined \def \binits#1{#1}\fi
\ifx \bauthor  \undefined \def \bauthor#1{#1}\fi
\ifx \batitle  \undefined \def \batitle#1{#1}\fi
\ifx \bjtitle  \undefined \def \bjtitle#1{#1}\fi
\ifx \bvolume  \undefined \def \bvolume#1{\textbf{#1}}\fi
\ifx \byear  \undefined \def \byear#1{#1}\fi
\ifx \bissue  \undefined \def \bissue#1{#1}\fi
\ifx \bfpage  \undefined \def \bfpage#1{#1}\fi
\ifx \blpage  \undefined \def \blpage #1{#1}\fi
\ifx \burl  \undefined \def \burl#1{\textsf{#1}}\fi
\ifx \doiurl  \undefined \def \doiurl#1{\url{https://doi.org/#1}}\fi
\ifx \betal  \undefined \def \betal{\textit{et al.}}\fi
\ifx \binstitute  \undefined \def \binstitute#1{#1}\fi
\ifx \binstitutionaled  \undefined \def \binstitutionaled#1{#1}\fi
\ifx \bctitle  \undefined \def \bctitle#1{#1}\fi
\ifx \beditor  \undefined \def \beditor#1{#1}\fi
\ifx \bpublisher  \undefined \def \bpublisher#1{#1}\fi
\ifx \bbtitle  \undefined \def \bbtitle#1{#1}\fi
\ifx \bedition  \undefined \def \bedition#1{#1}\fi
\ifx \bseriesno  \undefined \def \bseriesno#1{#1}\fi
\ifx \blocation  \undefined \def \blocation#1{#1}\fi
\ifx \bsertitle  \undefined \def \bsertitle#1{#1}\fi
\ifx \bsnm \undefined \def \bsnm#1{#1}\fi
\ifx \bsuffix \undefined \def \bsuffix#1{#1}\fi
\ifx \bparticle \undefined \def \bparticle#1{#1}\fi
\ifx \barticle \undefined \def \barticle#1{#1}\fi
\bibcommenthead
\ifx \bconfdate \undefined \def \bconfdate #1{#1}\fi
\ifx \botherref \undefined \def \botherref #1{#1}\fi
\ifx \url \undefined \def \url#1{\textsf{#1}}\fi
\ifx \bchapter \undefined \def \bchapter#1{#1}\fi
\ifx \bbook \undefined \def \bbook#1{#1}\fi
\ifx \bcomment \undefined \def \bcomment#1{#1}\fi
\ifx \oauthor \undefined \def \oauthor#1{#1}\fi
\ifx \citeauthoryear \undefined \def \citeauthoryear#1{#1}\fi
\ifx \endbibitem  \undefined \def \endbibitem {}\fi
\ifx \bconflocation  \undefined \def \bconflocation#1{#1}\fi
\ifx \arxivurl  \undefined \def \arxivurl#1{\textsf{#1}}\fi
\csname PreBibitemsHook\endcsname

\bibitem[\protect\citeauthoryear{Hasan and Kane}{2010}]{r1}
\begin{barticle}
\bauthor{\bsnm{Hasan}, \binits{M.Z.}},
\bauthor{\bsnm{Kane}, \binits{C.L.}}:
\batitle{Colloquium: Topological insulators}.
\bjtitle{Rev. Mod. Phys.}
\bvolume{82},
\bfpage{3045}--\blpage{3067}
(\byear{2010})
\doiurl{10.1103/RevModPhys.82.3045}
\end{barticle}
\endbibitem

\bibitem[\protect\citeauthoryear{Qi and Zhang}{2011}]{r2}
\begin{barticle}
\bauthor{\bsnm{Qi}, \binits{X.-L.}},
\bauthor{\bsnm{Zhang}, \binits{S.-C.}}:
\batitle{Topological insulators and superconductors}.
\bjtitle{Rev. Mod. Phys.}
\bvolume{83},
\bfpage{1057}--\blpage{1110}
(\byear{2011})
\doiurl{10.1103/RevModPhys.83.1057}
\end{barticle}
\endbibitem

\bibitem[\protect\citeauthoryear{Bansil et~al.}{2016}]{r3}
\begin{barticle}
\bauthor{\bsnm{Bansil}, \binits{A.}},
\bauthor{\bsnm{Lin}, \binits{H.}},
\bauthor{\bsnm{Das}, \binits{T.}}:
\batitle{Colloquium: Topological band theory}.
\bjtitle{Rev. Mod. Phys.}
\bvolume{88},
\bfpage{021004}
(\byear{2016})
\doiurl{10.1103/RevModPhys.88.021004}
\end{barticle}
\endbibitem

\bibitem[\protect\citeauthoryear{Nagaosa et~al.}{2010}]{r4}
\begin{barticle}
\bauthor{\bsnm{Nagaosa}, \binits{N.}},
\bauthor{\bsnm{Sinova}, \binits{J.}},
\bauthor{\bsnm{Onoda}, \binits{S.}},
\bauthor{\bsnm{MacDonald}, \binits{A.H.}},
\bauthor{\bsnm{Ong}, \binits{N.P.}}:
\batitle{Anomalous hall effect}.
\bjtitle{Rev. Mod. Phys.}
\bvolume{82},
\bfpage{1539}--\blpage{1592}
(\byear{2010})
\doiurl{10.1103/RevModPhys.82.1539}
\end{barticle}
\endbibitem

\bibitem[\protect\citeauthoryear{Sinova et~al.}{2015}]{r5}
\begin{barticle}
\bauthor{\bsnm{Sinova}, \binits{J.}},
\bauthor{\bsnm{Valenzuela}, \binits{S.O.}},
\bauthor{\bsnm{Wunderlich}, \binits{J.}},
\bauthor{\bsnm{Back}, \binits{C.H.}},
\bauthor{\bsnm{Jungwirth}, \binits{T.}}:
\batitle{Spin hall effects}.
\bjtitle{Rev. Mod. Phys.}
\bvolume{87},
\bfpage{1213}--\blpage{1260}
(\byear{2015})
\doiurl{10.1103/RevModPhys.87.1213}
\end{barticle}
\endbibitem

\bibitem[\protect\citeauthoryear{Hirsch}{1999}]{r6}
\begin{barticle}
\bauthor{\bsnm{Hirsch}, \binits{J.E.}}:
\batitle{Spin hall effect}.
\bjtitle{Phys. Rev. Lett.}
\bvolume{83},
\bfpage{1834}--\blpage{1837}
(\byear{1999})
\doiurl{10.1103/PhysRevLett.83.1834}
\end{barticle}
\endbibitem

\bibitem[\protect\citeauthoryear{Cinal et~al.}{1994}]{r7}
\begin{barticle}
\bauthor{\bsnm{Cinal}, \binits{M.}},
\bauthor{\bsnm{Edwards}, \binits{D.M.}},
\bauthor{\bsnm{Mathon}, \binits{J.}}:
\batitle{Magnetocrystalline anisotropy in ferromagnetic films}.
\bjtitle{Phys. Rev. B}
\bvolume{50},
\bfpage{3754}--\blpage{3760}
(\byear{1994})
\doiurl{10.1103/PhysRevB.50.3754}
\end{barticle}
\endbibitem

\bibitem[\protect\citeauthoryear{Haldane}{1988}]{r8}
\begin{barticle}
\bauthor{\bsnm{Haldane}, \binits{F.D.M.}}:
\batitle{Model for a quantum hall effect without landau levels: Condensed-matter realization of the "parity anomaly"}.
\bjtitle{Phys. Rev. Lett.}
\bvolume{61},
\bfpage{2015}--\blpage{2018}
(\byear{1988})
\doiurl{10.1103/PhysRevLett.61.2015}
\end{barticle}
\endbibitem

\bibitem[\protect\citeauthoryear{Chang et~al.}{2023}]{r9}
\begin{barticle}
\bauthor{\bsnm{Chang}, \binits{C.-Z.}},
\bauthor{\bsnm{Liu}, \binits{C.-X.}},
\bauthor{\bsnm{MacDonald}, \binits{A.H.}}:
\batitle{Colloquium: Quantum anomalous hall effect}.
\bjtitle{Rev. Mod. Phys.}
\bvolume{95},
\bfpage{011002}
(\byear{2023})
\doiurl{10.1103/RevModPhys.95.011002}
\end{barticle}
\endbibitem

\bibitem[\protect\citeauthoryear{Guo et~al.}{2023}]{r10}
\begin{barticle}
\bauthor{\bsnm{Guo}, \binits{P.-J.}},
\bauthor{\bsnm{Liu}, \binits{Z.-X.}},
\bauthor{\bsnm{Lu}, \binits{Z.-Y.}}:
\batitle{Quantum anomalous hall effect in collinear antiferromagnetism}.
\bjtitle{npj Comput. Mater.}
\bvolume{9}(\bissue{1}),
\bfpage{70}
(\byear{2023})
\doiurl{10.1038/s41524-023-01025-4}
\end{barticle}
\endbibitem

\bibitem[\protect\citeauthoryear{Liu et~al.}{2008}]{liu-prl}
\begin{barticle}
\bauthor{\bsnm{Liu}, \binits{G.-Q.}},
\bauthor{\bsnm{Antonov}, \binits{V.N.}},
\bauthor{\bsnm{Jepsen}, \binits{O.}},
\bauthor{\bsnm{Andersen.}, \binits{O.K.}}:
\batitle{Coulomb-enhanced spin-orbit splitting: The missing piece in the {${\mathrm{Sr}}_{2}{\mathrm{RhO}}_{4}$} puzzle}.
\bjtitle{Phys. Rev. Lett.}
\bvolume{101},
\bfpage{026408}
(\byear{2008})
\doiurl{10.1103/PhysRevLett.101.026408}
\end{barticle}
\endbibitem

\bibitem[\protect\citeauthoryear{Li et~al.}{2022}]{r11}
\begin{barticle}
\bauthor{\bsnm{Li}, \binits{J.-Y.}},
\bauthor{\bsnm{Yao}, \binits{Q.-S.}},
\bauthor{\bsnm{Wu}, \binits{L.}},
\bauthor{\bsnm{Hu}, \binits{Z.-X.}},
\bauthor{\bsnm{Gao}, \binits{B.-Y.}},
\bauthor{\bsnm{Wan}, \binits{X.-G.}},
\bauthor{\bsnm{Liu}, \binits{Q.-H.}}:
\batitle{Designing light-element materials with large effective spin-orbit coupling}.
\bjtitle{Nat. Commun.}
\bvolume{13}(\bissue{1}),
\bfpage{919}
(\byear{2022})
\doiurl{10.1038/s41467-022-28534-y}
\end{barticle}
\endbibitem

\bibitem[\protect\citeauthoryear{Šmejkal et~al.}{2022a}]{r12}
\begin{barticle}
\bauthor{\bsnm{Šmejkal}, \binits{L.}},
\bauthor{\bsnm{Sinova}, \binits{J.}},
\bauthor{\bsnm{Jungwirth}, \binits{T.}}:
\batitle{Beyond conventional ferromagnetism and antiferromagnetism: A phase with nonrelativistic spin and crystal rotation symmetry}.
\bjtitle{Phys. Rev. X}
\bvolume{12},
\bfpage{031042}
(\byear{2022})
\doiurl{10.1103/PhysRevX.12.031042}
\end{barticle}
\endbibitem

\bibitem[\protect\citeauthoryear{Šmejkal et~al.}{2022b}]{r13}
\begin{barticle}
\bauthor{\bsnm{Šmejkal}, \binits{L.}},
\bauthor{\bsnm{Sinova}, \binits{J.}},
\bauthor{\bsnm{Jungwirth}, \binits{T.}}:
\batitle{Emerging research landscape of altermagnetism}.
\bjtitle{Phys. Rev. X}
\bvolume{12},
\bfpage{040501}
(\byear{2022})
\doiurl{10.1103/PhysRevX.12.040501}
\end{barticle}
\endbibitem

\bibitem[\protect\citeauthoryear{Gao et~al.}{2025}]{r14}
\begin{barticle}
\bauthor{\bsnm{Gao}, \binits{Z.-F.}},
\bauthor{\bsnm{Qu}, \binits{S.}},
\bauthor{\bsnm{Zeng}, \binits{B.}},
\bauthor{\bsnm{Liu}, \binits{Y.}},
\bauthor{\bsnm{Wen}, \binits{J.-R.}},
\bauthor{\bsnm{Sun}, \binits{H.}},
\bauthor{\bsnm{Guo}, \binits{P.-J.}},
\bauthor{\bsnm{Lu}, \binits{Z.-Y.}}:
\batitle{Ai-accelerated discovery of altermagnetic materials}.
\bjtitle{National Science Review}
\bvolume{12}(\bissue{4}),
\bfpage{066}
(\byear{2025})
\doiurl{10.1093/nsr/nwaf066}
{\href{https://arxiv.org/abs/https://academic.oup.com/nsr/article-pdf/12/4/nwaf066/62054665/nwaf066.pdf}{{https://academic.oup.com/nsr/article-pdf/12/4/nwaf066/62054665/nwaf066.pdf}}}
\end{barticle}
\endbibitem

\bibitem[\protect\citeauthoryear{Ma et~al.}{2021}]{r15}
\begin{barticle}
\bauthor{\bsnm{Ma}, \binits{H.-Y.}},
\bauthor{\bsnm{Hu}, \binits{M.-L.}},
\bauthor{\bsnm{Li}, \binits{N.-N.}},
\bauthor{\bsnm{Liu}, \binits{J.-P.}},
\bauthor{\bsnm{Yao}, \binits{W.}},
\bauthor{\bsnm{Jia}, \binits{J.-F.}},
\bauthor{\bsnm{Liu}, \binits{J.-W.}}:
\batitle{Multifunctional antiferromagnetic materials with giant piezomagnetism and noncollinear spin current}.
\bjtitle{Nat. Commun.}
\bvolume{12}(\bissue{1}),
\bfpage{2846}
(\byear{2021})
\doiurl{10.1038/s41467-021-23127-7}
\end{barticle}
\endbibitem

\bibitem[\protect\citeauthoryear{Šmejkal et~al.}{2022}]{r16}
\begin{barticle}
\bauthor{\bsnm{Šmejkal}, \binits{L.}},
\bauthor{\bsnm{Hellenes}, \binits{A.B.}},
\bauthor{\bsnm{Gonz\'alez-Hern\'andez}, \binits{R.}},
\bauthor{\bsnm{Sinova}, \binits{J.}},
\bauthor{\bsnm{Jungwirth}, \binits{T.}}:
\batitle{Giant and tunneling magnetoresistance in unconventional collinear antiferromagnets with nonrelativistic spin-momentum coupling}.
\bjtitle{Phys. Rev. X}
\bvolume{12},
\bfpage{011028}
(\byear{2022})
\doiurl{10.1103/PhysRevX.12.011028}
\end{barticle}
\endbibitem

\bibitem[\protect\citeauthoryear{Gonz\'alez-Hern\'andez et~al.}{2021}]{r17}
\begin{barticle}
\bauthor{\bsnm{Gonz\'alez-Hern\'andez}, \binits{R.}},
\bauthor{\bsnm{Šmejkal}, \binits{L.}},
\bauthor{\bsnm{V\'yborn\'y}, \binits{K.}},
\bauthor{\bsnm{Yahagi}, \binits{Y.}},
\bauthor{\bsnm{Sinova}, \binits{J.}},
\bauthor{\bsnm{Jungwirth}, \binits{T.}},
\bauthor{\bsnm{Železn\'y}, \binits{J.}}:
\batitle{Efficient electrical spin splitter based on nonrelativistic collinear antiferromagnetism}.
\bjtitle{Phys. Rev. Lett.}
\bvolume{126},
\bfpage{127701}
(\byear{2021})
\doiurl{10.1103/PhysRevLett.126.127701}
\end{barticle}
\endbibitem

\bibitem[\protect\citeauthoryear{Bai et~al.}{2022}]{r18}
\begin{barticle}
\bauthor{\bsnm{Bai}, \binits{H.}},
\bauthor{\bsnm{Han}, \binits{L.}},
\bauthor{\bsnm{Feng}, \binits{X.Y.}},
\bauthor{\bsnm{Zhou}, \binits{Y.J.}},
\bauthor{\bsnm{Su}, \binits{R.X.}},
\bauthor{\bsnm{Wang}, \binits{Q.}},
\bauthor{\bsnm{Liao}, \binits{L.Y.}},
\bauthor{\bsnm{Zhu}, \binits{W.X.}},
\bauthor{\bsnm{Chen}, \binits{X.Z.}},
\bauthor{\bsnm{Pan}, \binits{F.}},
\bauthor{\bsnm{Fan}, \binits{X.L.}},
\bauthor{\bsnm{Song}, \binits{C.}}:
\batitle{Observation of spin splitting torque in a collinear antiferromagnet {${\mathrm{RuO}}_{2}$}}.
\bjtitle{Phys. Rev. Lett.}
\bvolume{128},
\bfpage{197202}
(\byear{2022})
\doiurl{10.1103/PhysRevLett.128.197202}
\end{barticle}
\endbibitem

\bibitem[\protect\citeauthoryear{Karube et~al.}{2022}]{r19}
\begin{barticle}
\bauthor{\bsnm{Karube}, \binits{S.}},
\bauthor{\bsnm{Tanaka}, \binits{T.}},
\bauthor{\bsnm{Sugawara}, \binits{D.}},
\bauthor{\bsnm{Kadoguchi}, \binits{N.}},
\bauthor{\bsnm{Kohda}, \binits{M.}},
\bauthor{\bsnm{Nitta}, \binits{J.}}:
\batitle{Observation of spin-splitter torque in collinear antiferromagnetic {${\mathrm{RuO}}_{2}$}}.
\bjtitle{Phys. Rev. Lett.}
\bvolume{129},
\bfpage{137201}
(\byear{2022})
\doiurl{10.1103/PhysRevLett.129.137201}
\end{barticle}
\endbibitem

\bibitem[\protect\citeauthoryear{Zhu et~al.}{2023}]{r20}
\begin{barticle}
\bauthor{\bsnm{Zhu}, \binits{D.}},
\bauthor{\bsnm{Zhuang}, \binits{Z.-Y.}},
\bauthor{\bsnm{Wu}, \binits{Z.-G.}},
\bauthor{\bsnm{Yan}, \binits{Z.-B.}}:
\batitle{Topological superconductivity in two-dimensional altermagnetic metals}.
\bjtitle{Phys. Rev. B}
\bvolume{108},
\bfpage{184505}
(\byear{2023})
\doiurl{10.1103/PhysRevB.108.184505}
\end{barticle}
\endbibitem

\bibitem[\protect\citeauthoryear{Guo et~al.}{2023}]{r21}
\begin{botherref}
\oauthor{\bsnm{Guo}, \binits{P.-J.}},
\oauthor{\bsnm{Gu}, \binits{Y.-H.}},
\oauthor{\bsnm{Gao}, \binits{Z.-F.}},
\oauthor{\bsnm{Lu}, \binits{Z.-Y.}}:
Altermagnetic ferroelectric {$\mathrm{LiFe}_2\mathrm{F}_6$} and spin-triplet excitonic insulator phase
(2023)
\end{botherref}
\endbibitem

\bibitem[\protect\citeauthoryear{Zhou et~al.}{2024}]{r22}
\begin{barticle}
\bauthor{\bsnm{Zhou}, \binits{X.-D.}},
\bauthor{\bsnm{Feng}, \binits{W.-X.}},
\bauthor{\bsnm{Zhang}, \binits{R.-W.}},
\bauthor{\bsnm{Šmejkal}, \binits{L.}},
\bauthor{\bsnm{Sinova}, \binits{J.}},
\bauthor{\bsnm{Mokrousov}, \binits{Y.}},
\bauthor{\bsnm{Yao}, \binits{Y.-G.}}:
\batitle{Crystal thermal transport in altermagnetic {${\mathrm{RuO}}_{2}$}}.
\bjtitle{Phys. Rev. Lett.}
\bvolume{132},
\bfpage{056701}
(\byear{2024})
\doiurl{10.1103/PhysRevLett.132.056701}
\end{barticle}
\endbibitem

\bibitem[\protect\citeauthoryear{Šmejkal et~al.}{2020}]{r23}
\begin{barticle}
\bauthor{\bsnm{Šmejkal}, \binits{L.}},
\bauthor{\bsnm{González-Hernández}, \binits{R.}},
\bauthor{\bsnm{Jungwirth}, \binits{T.}},
\bauthor{\bsnm{Sinova}, \binits{J.}}:
\batitle{Crystal time-reversal symmetry breaking and spontaneous hall effect in collinear antiferromagnets}.
\bjtitle{Sci. Adv.}
\bvolume{6}(\bissue{23}),
\bfpage{8809}
(\byear{2020})
\doiurl{10.1126/sciadv.aaz8809}
\end{barticle}
\endbibitem

\bibitem[\protect\citeauthoryear{Zhou et~al.}{2021}]{r24}
\begin{barticle}
\bauthor{\bsnm{Zhou}, \binits{X.-D.}},
\bauthor{\bsnm{Feng}, \binits{W.-X.}},
\bauthor{\bsnm{Yang}, \binits{X.-X.}},
\bauthor{\bsnm{Guo}, \binits{G.-Y.}},
\bauthor{\bsnm{Yao}, \binits{Y.-G.}}:
\batitle{Crystal chirality magneto-optical effects in collinear antiferromagnets}.
\bjtitle{Phys. Rev. B}
\bvolume{104},
\bfpage{024401}
(\byear{2021})
\doiurl{10.1103/PhysRevB.104.024401}
\end{barticle}
\endbibitem

\bibitem[\protect\citeauthoryear{Hou et~al.}{2023}]{r25}
\begin{barticle}
\bauthor{\bsnm{Hou}, \binits{X.-Y.}},
\bauthor{\bsnm{Yang}, \binits{H.-C.}},
\bauthor{\bsnm{Liu}, \binits{Z.-X.}},
\bauthor{\bsnm{Guo}, \binits{P.-J.}},
\bauthor{\bsnm{Lu}, \binits{Z.-Y.}}:
\batitle{Large intrinsic anomalous hall effect in both {${\mathrm{Nb}}_{2}{\mathrm{FeB}}_{2}$} and {${\mathrm{Ta}}_{2}{\mathrm{FeB}}_{2}$} with collinear antiferromagnetism}.
\bjtitle{Phys. Rev. B}
\bvolume{107},
\bfpage{161109}
(\byear{2023})
\doiurl{10.1103/PhysRevB.107.L161109}
\end{barticle}
\endbibitem

\bibitem[\protect\citeauthoryear{Guo et~al.}{2021}]{r26}
\begin{barticle}
\bauthor{\bsnm{Guo}, \binits{P.-J.}},
\bauthor{\bsnm{Wei}, \binits{Y.-W.}},
\bauthor{\bsnm{Liu}, \binits{K.}},
\bauthor{\bsnm{Liu}, \binits{Z.-X.}},
\bauthor{\bsnm{Lu}, \binits{Z.-Y.}}:
\batitle{Eightfold degenerate fermions in two dimensions}.
\bjtitle{Phys. Rev. Lett.}
\bvolume{127},
\bfpage{176401}
(\byear{2021})
\doiurl{10.1103/PhysRevLett.127.176401}
\end{barticle}
\endbibitem

\bibitem[\protect\citeauthoryear{Xiao et~al.}{2024}]{r27}
\begin{barticle}
\bauthor{\bsnm{Xiao}, \binits{Z.-Y.}},
\bauthor{\bsnm{Zhao}, \binits{J.-Z.}},
\bauthor{\bsnm{Li}, \binits{Y.-Q.}},
\bauthor{\bsnm{Shindou}, \binits{R.}},
\bauthor{\bsnm{Song}, \binits{Z.-D.}}:
\batitle{Spin space groups: Full classification and applications}.
\bjtitle{Phys. Rev. X}
\bvolume{14},
\bfpage{031037}
(\byear{2024})
\doiurl{10.1103/PhysRevX.14.031037}
\end{barticle}
\endbibitem

\bibitem[\protect\citeauthoryear{Jiang et~al.}{2024}]{r28}
\begin{barticle}
\bauthor{\bsnm{Jiang}, \binits{Y.}},
\bauthor{\bsnm{Song}, \binits{Z.-Y.}},
\bauthor{\bsnm{Zhu}, \binits{T.-N.}},
\bauthor{\bsnm{Fang}, \binits{Z.}},
\bauthor{\bsnm{Weng}, \binits{H.-M.}},
\bauthor{\bsnm{Liu}, \binits{Z.-X.}},
\bauthor{\bsnm{Yang}, \binits{J.}},
\bauthor{\bsnm{Fang}, \binits{C.}}:
\batitle{Enumeration of spin-space groups: Toward a complete description of symmetries of magnetic orders}.
\bjtitle{Phys. Rev. X}
\bvolume{14},
\bfpage{031039}
(\byear{2024})
\doiurl{10.1103/PhysRevX.14.031039}
\end{barticle}
\endbibitem

\bibitem[\protect\citeauthoryear{Chen et~al.}{2025}]{r29}
\begin{barticle}
\bauthor{\bsnm{Chen}, \binits{X.}},
\bauthor{\bsnm{Liu}, \binits{Y.}},
\bauthor{\bsnm{Liu}, \binits{P.}},
\bauthor{\bsnm{Yu}, \binits{Y.}},
\bauthor{\bsnm{Ren}, \binits{J.}},
\bauthor{\bsnm{Li}, \binits{J.}},
\bauthor{\bsnm{Zhang}, \binits{A.}},
\bauthor{\bsnm{Liu}, \binits{Q.}}:
\batitle{Unconventional magnons in collinear magnets dictated by spin space groups}.
\bjtitle{Nature}
\bvolume{640}(\bissue{8058}),
\bfpage{349}--\blpage{354}
(\byear{2025})
\doiurl{10.1038/s41586-025-08715-7}
\end{barticle}
\endbibitem

\bibitem[\protect\citeauthoryear{Zemva et~al.}{1995}]{r30}
\begin{barticle}
\bauthor{\bsnm{Zemva}, \binits{B.}},
\bauthor{\bsnm{Lutar}, \binits{K.}},
\bauthor{\bsnm{Chacon}, \binits{L.}},
\bauthor{\bsnm{Fele-Beuermann}, \binits{M.}},
\bauthor{\bsnm{Allman}, \binits{J.}},
\bauthor{\bsnm{Shen}, \binits{C.}},
\bauthor{\bsnm{Bartlett}, \binits{N.}}:
\batitle{Thermodynamically unstable fluorides of nickel: {$\mathrm{NiF}_4$} and {$\mathrm{NiF}_3$} syntheses and some properties}.
\bjtitle{J. Am. Chem. Soc.}
\bvolume{117}(\bissue{40}),
\bfpage{10025}--\blpage{10034}
(\byear{1995})
\doiurl{10.1021/ja00145a013}
\end{barticle}
\endbibitem

\bibitem[\protect\citeauthoryear{Mattsson and Paulus}{2019}]{r31}
\begin{barticle}
\bauthor{\bsnm{Mattsson}, \binits{S.}},
\bauthor{\bsnm{Paulus}, \binits{B.}}:
\batitle{Density functional theory calculations of structural, electronic, and magnetic properties of the {$3\mathrm{d}$} metal trifluorides {$\mathrm{MF}_3$} ({$\mathrm{M}=\mathrm{Ti}-\mathrm{Ni}$}) in the solid state}.
\bjtitle{J. Comput. Chem.}
\bvolume{40}(\bissue{11}),
\bfpage{1190}--\blpage{1197}
(\byear{2019})
\doiurl{10.1002/jcc.25777}
\end{barticle}
\endbibitem

\bibitem[\protect\citeauthoryear{Goodenough}{2008}]{r32}
\begin{barticle}
\bauthor{\bsnm{Goodenough}, \binits{J.B.}}:
\batitle{{G}oodenough-{K}anamori rule}.
\bjtitle{Scholarpedia}
\bvolume{3}(\bissue{10}),
\bfpage{7382}
(\byear{2008})
\doiurl{10.4249/scholarpedia.7382} .
\bcomment{revision \#122456}
\end{barticle}
\endbibitem

\bibitem[\protect\citeauthoryear{}{}]{r36}
\begin{botherref}
Supplemental material
\end{botherref}
\endbibitem

\bibitem[\protect\citeauthoryear{Wang et~al.}{2022}]{r33}
\begin{barticle}
\bauthor{\bsnm{Wang}, \binits{K.-L.}},
\bauthor{\bsnm{Huang}, \binits{W.}},
\bauthor{\bsnm{Cao}, \binits{Q.-H.}},
\bauthor{\bsnm{Zhao}, \binits{Y.-J.}},
\bauthor{\bsnm{Sun}, \binits{X.-J.}},
\bauthor{\bsnm{Ding}, \binits{R.}},
\bauthor{\bsnm{Lin}, \binits{W.-W.}},
\bauthor{\bsnm{Liu}, \binits{E.-H.}},
\bauthor{\bsnm{Gao}, \binits{P.}}:
\batitle{Engineering {$\mathrm{NiF}_3/\mathrm{Ni}_2\mathrm{P}$} heterojunction as efficient electrocatalysts for urea oxidation and splitting}.
\bjtitle{Chem. Eng. J.}
\bvolume{427},
\bfpage{130865}
(\byear{2022})
\doiurl{10.1016/j.cej.2021.130865}
\end{barticle}
\endbibitem

\bibitem[\protect\citeauthoryear{Effenberger et~al.}{1981}]{r48}
\begin{barticle}
\bauthor{\bsnm{Effenberger}, \binits{H.}},
\bauthor{\bsnm{Mereiter}, \binits{K.}},
\bauthor{\bsnm{Zemann}, \binits{J.}}:
\batitle{Crystal structure refinements of magnesite, calcite, rhodochrosite, siderite, smithonite, and dolomite, with discussion of some aspects of the stereochemistry of calcite type carbonates}.
\bjtitle{Z. Kristallogr. Cryst. Mater.}
\bvolume{156}(\bissue{1-4}),
\bfpage{233}--\blpage{244}
(\byear{1981})
\end{barticle}
\endbibitem

\bibitem[\protect\citeauthoryear{Alikhanov}{1959}]{r49}
\begin{barticle}
\bauthor{\bsnm{Alikhanov}, \binits{R.A.}}:
\batitle{Neutron diffraction investigation of the antiferromagnetism of the carbonates of manganese and iron}.
\bjtitle{Sov. Phys. JETP-USSR}
\bvolume{9}(\bissue{6}),
\bfpage{1204}--\blpage{1208}
(\byear{1959})
\end{barticle}
\endbibitem

\bibitem[\protect\citeauthoryear{Hohenberg and Kohn}{1964}]{r37}
\begin{barticle}
\bauthor{\bsnm{Hohenberg}, \binits{P.}},
\bauthor{\bsnm{Kohn}, \binits{W.}}:
\batitle{Inhomogeneous electron gas}.
\bjtitle{Phys. Rev.}
\bvolume{136},
\bfpage{864}--\blpage{871}
(\byear{1964})
\doiurl{10.1103/PhysRev.136.B864}
\end{barticle}
\endbibitem

\bibitem[\protect\citeauthoryear{Kohn and Sham}{1965}]{r38}
\begin{barticle}
\bauthor{\bsnm{Kohn}, \binits{W.}},
\bauthor{\bsnm{Sham}, \binits{L.J.}}:
\batitle{Self-consistent equations including exchange and correlation effects}.
\bjtitle{Phys. Rev.}
\bvolume{140},
\bfpage{1133}--\blpage{1138}
(\byear{1965})
\doiurl{10.1103/PhysRev.140.A1133}
\end{barticle}
\endbibitem

\bibitem[\protect\citeauthoryear{Kresse and Furthmüller}{1996}]{r39}
\begin{barticle}
\bauthor{\bsnm{Kresse}, \binits{G.}},
\bauthor{\bsnm{Furthmüller}, \binits{J.}}:
\batitle{Efficiency of ab-initio total energy calculations for metals and semiconductors using a plane-wave basis set}.
\bjtitle{Comput. Mater. Sci.}
\bvolume{6}(\bissue{1}),
\bfpage{15}--\blpage{50}
(\byear{1996})
\doiurl{10.1016/0927-0256(96)00008-0}
\end{barticle}
\endbibitem

\bibitem[\protect\citeauthoryear{Kresse and Hafner}{1993}]{r40}
\begin{barticle}
\bauthor{\bsnm{Kresse}, \binits{G.}},
\bauthor{\bsnm{Hafner}, \binits{J.}}:
\batitle{Ab initio molecular dynamics for liquid metals}.
\bjtitle{Phys. Rev. B}
\bvolume{47},
\bfpage{558}--\blpage{561}
(\byear{1993})
\doiurl{10.1103/PhysRevB.47.558}
\end{barticle}
\endbibitem

\bibitem[\protect\citeauthoryear{Kresse and Furthm\"uller}{1996}]{r41}
\begin{barticle}
\bauthor{\bsnm{Kresse}, \binits{G.}},
\bauthor{\bsnm{Furthm\"uller}, \binits{J.}}:
\batitle{Efficient iterative schemes for ab initio total-energy calculations using a plane-wave basis set}.
\bjtitle{Phys. Rev. B}
\bvolume{54},
\bfpage{11169}--\blpage{11186}
(\byear{1996})
\doiurl{10.1103/PhysRevB.54.11169}
\end{barticle}
\endbibitem

\bibitem[\protect\citeauthoryear{Giannozzi et~al.}{2020}]{r52}
\begin{barticle}
\bauthor{\bsnm{Giannozzi}, \binits{P.}},
\bauthor{\bsnm{Baseggio}, \binits{O.}},
\bauthor{\bsnm{Bonfà}, \binits{P.}},
\bauthor{\bsnm{Brunato}, \binits{D.}},
\bauthor{\bsnm{Car}, \binits{R.}},
\bauthor{\bsnm{Carnimeo}, \binits{I.}},
\bauthor{\bsnm{Cavazzoni}, \binits{C.}},
\bauthor{\bsnm{Gironcoli}, \binits{S.}},
\bauthor{\bsnm{Delugas}, \binits{P.}},
\bauthor{\bsnm{Ferrari~Ruffino}, \binits{F.}},
\bauthor{\bsnm{Ferretti}, \binits{A.}},
\bauthor{\bsnm{Marzari}, \binits{N.}},
\bauthor{\bsnm{Timrov}, \binits{I.}},
\bauthor{\bsnm{Urru}, \binits{A.}},
\bauthor{\bsnm{Baroni}, \binits{S.}}:
\batitle{Quantum espresso toward the exascale}.
\bjtitle{J. Chem. Phys.}
\bvolume{152}(\bissue{15}),
\bfpage{154105}
(\byear{2020})
\doiurl{10.1063/5.0005082}
\end{barticle}
\endbibitem

\bibitem[\protect\citeauthoryear{Perdew et~al.}{1996}]{r42}
\begin{barticle}
\bauthor{\bsnm{Perdew}, \binits{J.P.}},
\bauthor{\bsnm{Burke}, \binits{K.}},
\bauthor{\bsnm{Ernzerhof}, \binits{M.}}:
\batitle{Generalized gradient approximation made simple}.
\bjtitle{Phys. Rev. Lett.}
\bvolume{77},
\bfpage{3865}--\blpage{3868}
(\byear{1996})
\doiurl{10.1103/PhysRevLett.77.3865}
\end{barticle}
\endbibitem

\bibitem[\protect\citeauthoryear{Sun et~al.}{2015}]{r51}
\begin{barticle}
\bauthor{\bsnm{Sun}, \binits{J.-W.}},
\bauthor{\bsnm{Ruzsinszky}, \binits{A.}},
\bauthor{\bsnm{Perdew}, \binits{J.P.}}:
\batitle{Strongly constrained and appropriately normed semilocal density functional}.
\bjtitle{Phys. Rev. Lett.}
\bvolume{115},
\bfpage{036402}
(\byear{2015})
\doiurl{10.1103/PhysRevLett.115.036402}
\end{barticle}
\endbibitem

\bibitem[\protect\citeauthoryear{Bl\"ochl}{1994}]{r43}
\begin{barticle}
\bauthor{\bsnm{Bl\"ochl}, \binits{P.E.}}:
\batitle{Projector augmented-wave method}.
\bjtitle{Phys. Rev. B}
\bvolume{50},
\bfpage{17953}--\blpage{17979}
(\byear{1994})
\doiurl{10.1103/PhysRevB.50.17953}
\end{barticle}
\endbibitem

\bibitem[\protect\citeauthoryear{Kresse and Joubert}{1999}]{r44}
\begin{barticle}
\bauthor{\bsnm{Kresse}, \binits{G.}},
\bauthor{\bsnm{Joubert}, \binits{D.}}:
\batitle{From ultrasoft pseudopotentials to the projector augmented-wave method}.
\bjtitle{Phys. Rev. B}
\bvolume{59},
\bfpage{1758}--\blpage{1775}
(\byear{1999})
\doiurl{10.1103/PhysRevB.59.1758}
\end{barticle}
\endbibitem

\bibitem[\protect\citeauthoryear{Anisimov et~al.}{1991}]{r45}
\begin{barticle}
\bauthor{\bsnm{Anisimov}, \binits{V.I.}},
\bauthor{\bsnm{Zaanen}, \binits{J.}},
\bauthor{\bsnm{Andersen}, \binits{O.K.}}:
\batitle{Band theory and mott insulators: Hubbard {$\mathrm{U}$} instead of stoner {$\mathrm{I}$}}.
\bjtitle{Phys. Rev. B}
\bvolume{44},
\bfpage{943}--\blpage{954}
(\byear{1991})
\doiurl{10.1103/PhysRevB.44.943}
\end{barticle}
\endbibitem

\bibitem[\protect\citeauthoryear{Liechtenstein et~al.}{1995}]{r50}
\begin{barticle}
\bauthor{\bsnm{Liechtenstein}, \binits{A.I.}},
\bauthor{\bsnm{Anisimov}, \binits{V.I.}},
\bauthor{\bsnm{Zaanen}, \binits{J.}}:
\batitle{Density-functional theory and strong interactions: Orbital ordering in mott-hubbard insulators}.
\bjtitle{Phys. Rev. B}
\bvolume{52},
\bfpage{5467}--\blpage{5470}
(\byear{1995})
\doiurl{10.1103/PhysRevB.52.R5467}
\end{barticle}
\endbibitem

\bibitem[\protect\citeauthoryear{Dudarev et~al.}{1998}]{r46}
\begin{barticle}
\bauthor{\bsnm{Dudarev}, \binits{S.L.}},
\bauthor{\bsnm{Botton}, \binits{G.A.}},
\bauthor{\bsnm{Savrasov}, \binits{S.Y.}},
\bauthor{\bsnm{Humphreys}, \binits{C.J.}},
\bauthor{\bsnm{Sutton}, \binits{A.P.}}:
\batitle{Electron-energy-loss spectra and the structural stability of nickel oxide: An {$\mathrm{LSDA}+\mathrm{U}$} study}.
\bjtitle{Phys. Rev. B}
\bvolume{57},
\bfpage{1505}--\blpage{1509}
(\byear{1998})
\doiurl{10.1103/PhysRevB.57.1505}
\end{barticle}
\endbibitem

\bibitem[\protect\citeauthoryear{Zhou et~al.}{2004}]{r47}
\begin{barticle}
\bauthor{\bsnm{Zhou}, \binits{F.}},
\bauthor{\bsnm{Cococcioni}, \binits{M.}},
\bauthor{\bsnm{Marianetti}, \binits{C.A.}},
\bauthor{\bsnm{Morgan}, \binits{D.}},
\bauthor{\bsnm{Ceder}, \binits{G.}}:
\batitle{First-principles prediction of redox potentials in transition-metal compounds with {$\mathrm{LDA}+U$}}.
\bjtitle{Phys. Rev. B}
\bvolume{70},
\bfpage{235121}
(\byear{2004})
\doiurl{10.1103/PhysRevB.70.235121}
\end{barticle}
\endbibitem

\bibitem[\protect\citeauthoryear{Krukau et~al.}{2006}]{r53}
\begin{barticle}
\bauthor{\bsnm{Krukau}, \binits{A.V.}},
\bauthor{\bsnm{Vydrov}, \binits{O.A.}},
\bauthor{\bsnm{Izmaylov}, \binits{A.F.}},
\bauthor{\bsnm{Scuseria}, \binits{G.E.}}:
\batitle{Influence of the exchange screening parameter on the performance of screened hybrid functionals}.
\bjtitle{J. Chem. Phys.}
\bvolume{125}(\bissue{22}),
\bfpage{224106}
(\byear{2006})
\doiurl{10.1063/1.2404663}
\end{barticle}
\endbibitem

\bibitem[\protect\citeauthoryear{Haule}{2007}]{r54}
\begin{barticle}
\bauthor{\bsnm{Haule}, \binits{K.}}:
\batitle{Quantum monte carlo impurity solver for cluster dynamical mean-field theory and electronic structure calculations with adjustable cluster base}.
\bjtitle{Phys. Rev. B}
\bvolume{75},
\bfpage{155113}
(\byear{2007})
\doiurl{10.1103/PhysRevB.75.155113}
\end{barticle}
\endbibitem

\bibitem[\protect\citeauthoryear{Haule et~al.}{2010}]{r55}
\begin{barticle}
\bauthor{\bsnm{Haule}, \binits{K.}},
\bauthor{\bsnm{Yee}, \binits{C.-H.}},
\bauthor{\bsnm{Kim}, \binits{K.}}:
\batitle{Dynamical mean-field theory within the full-potential methods: Electronic structure of {${\text{CeIrIn}}_{5}$}, {${\text{CeCoIn}}_{5}$}, and {${\text{CeRhIn}}_{5}$}}.
\bjtitle{Phys. Rev. B}
\bvolume{81},
\bfpage{195107}
(\byear{2010})
\doiurl{10.1103/PhysRevB.81.195107}
\end{barticle}
\endbibitem

\bibitem[\protect\citeauthoryear{Haule}{2015}]{r56}
\begin{barticle}
\bauthor{\bsnm{Haule}, \binits{K.}}:
\batitle{Exact double counting in combining the dynamical mean field theory and the density functional theory}.
\bjtitle{Phys. Rev. Lett.}
\bvolume{115},
\bfpage{196403}
(\byear{2015})
\doiurl{10.1103/PhysRevLett.115.196403}
\end{barticle}
\endbibitem

\bibitem[\protect\citeauthoryear{Lee et~al.}{2018}]{r35}
\begin{barticle}
\bauthor{\bsnm{Lee}, \binits{S.}},
\bauthor{\bsnm{Torii}, \binits{S.}},
\bauthor{\bsnm{Ishikawa}, \binits{Y.}},
\bauthor{\bsnm{Yonemura}, \binits{M.}},
\bauthor{\bsnm{Moyoshi}, \binits{T.}},
\bauthor{\bsnm{Kamiyama}, \binits{T.}}:
\batitle{Weak-ferromagnetism of {$\mathrm{CoF}_3$} and {$\mathrm{FeF}_3$}}.
\bjtitle{Physica B Condens. Matter}
\bvolume{551},
\bfpage{94}--\blpage{97}
(\byear{2018})
\doiurl{10.1016/j.physb.2017.11.082} .
\bcomment{The 11th International Conference on Neutron Scattering (ICNS 2017)}
\end{barticle}
\endbibitem

\end{thebibliography}

\section*{Acknowledgements}

We thank Z.-X. Liu, X.-H. Kong, and C. Wang for valuable discussions.
This work was financially supported by the National Key R$\&$D Program of China (Grant No.2024YFA1408601), the National Natural Science Foundation of China (Grant No.12434009, No.12204533, No.62206299 and No.12174443) and the Beijing Natural Science Foundation (Grant No.Z200005). Computational resources have been provided by the Physical Laboratory of High Performance Computing at Renmin University of China.

\section*{Author contributions}
P.-J.G. and Z.-Y.L. contributed to the ideation and design of the research; S.Q. and Z.-F. O. performed the calculations; P.-J.G. and Z.-Y.L. performed symmetry analysis, wrote and edited the paper; all authors contributed to the research discussions.

\section*{Competing interests}
The authors declare no competing interests.
\section*{Correspondence}

\begin{appendices}

\section{Section title of first appendix}\label{secA1}

\subsection{Magnetocrystalline anisotropy}
\begin{figure*}[htbp]
\centering
\includegraphics[width=0.75\textwidth]{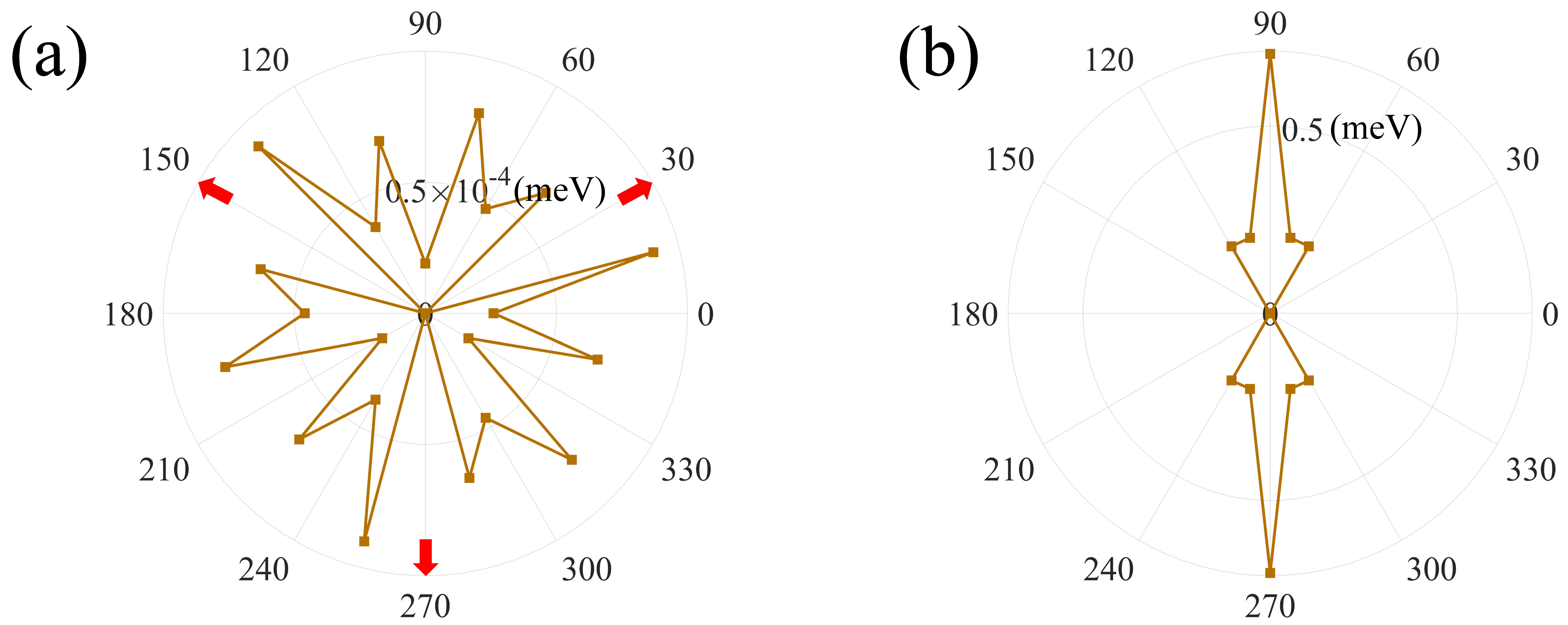}
\caption{Magnetocrystalline anisotropy of $\rm NiF_3$ for the $ x - y$ plane (a) and the $ y - z$ plane (b). In (a), the $ x$ and $ y$ axes represent 0 degrees and 90 degrees, respectively. In (b), the $ y$ and $ z$ axes represent 0 degrees and 90 degrees, respectively. The red arrows represent the direction of easy magnetization axes.
}\label{fig:5}
\end{figure*}

In the relativistic case, the coupling of spin and lattice causes crystal material to change from spin group symmetry to magnetic group symmetry, and the direction of easy magnetization axis determines the magnetic group symmetry of the crystal material. For altermagnetic $\rm NiF_3$, by calculating the magnetocrystalline anisotropy energy in both $ x - y$ and $ y - z$ planes, we find the easy magnetization axis is along 30 degrees, 150 degrees, and 270 degrees directions in the $ x - y$ plane, deriving from the $\rm C_3$ symmetry (Fig.\ref{fig:5} (a) and (b)). Furthermore, our analysis shows that altermagnetic $\rm NiF_3$ has $\rm I$, $\rm TC_2^1 t$ and $\rm TM_1 t$ symmetries. Thus, when soc is considered, the quadruple degenerate band on the $\rm \Gamma-T$ axis will split into four non-degenerate bands. Different from $\rm NiF_3$, the easy magnetization axis of $\rm FeCO_3$ and $\rm CoF_3$ have been confirmed by neutron scattering experiments and are along the z direction of the crystallographic unit cell \cite{r49,r35}, which make the symmetry of $\rm FeCO_3$ and $\rm CoF_3$ being double point group $\rm D_{3d}$.

\subsection{The electronic structures of the other five materials}

\begin{figure*}[htbp]
\centering
\includegraphics[width=0.80\textwidth]{figure/fs2.jpg}
\caption{The electronic band structures of $\rm AF_3$ (A=V, Cr, Fe, Co). (a) and (e) are the electronic band structures of $\rm VF_3$ without and with SOC under the correlation interaction ($\rm U=$ 3.3 $\rm eV$), respectively. (b) and (f) are the electronic band structures of $\rm CrF_3$ without and with SOC under the correlation interaction ($\rm U=$ 3.4 $\rm eV$), respectively. (c) and (g) are the electronic band structures of $\rm FeF_3$ without and with SOC under the correlation interaction ($\rm U=$ 4.9 $\rm eV$), respectively. (d) and (h) are the electronic band structures of $\rm CoF_3$ without and with SOC under the correlation interaction ($\rm U=$ 4.0 $\rm eV$), respectively.
}\label{fig:6}
\end{figure*}

\begin{table*}[htbp]
\centering
\caption{\label{tab:table1}
The relationship between valence electron occupation, crystal symmetry and SOC of altermagnetic materials. The VE, QD, DD, and EO represent valance electrons, quadruple degeneracy, double degeneracy and electron occupation, respectively.}
\begin{tabular}{|c|c|c|c|c|c|c|}
\hline
Compound & $\rm VF_3$ & $\rm CrF_3$ & $\rm FeF_3$ & $\rm CoF_3$ & $\rm NiF_3$ & $\rm FeCO_3$ \\ \hline
VE & 64 & 66 & 70 & 72 & 74 & 72\\ \hline
QD & 11 & 11 & 12 & 12 & 13 & 13\\ \hline
DD & 10 & 11 & 11 & 12 & 12 & 11\\ \hline
EO & Full & Full & Full & Full & Half & Half\\ \hline
SOC & Weak & Weak & Weak & Weak & Strong & Strong\\\hline
\end{tabular}
\end{table*}

Since $\rm AF_3$ (A=V, Cr, Fe, Co) and $\rm NiF_3$ have the same crystal structure and magnetic structure, $\rm AF_3$ (A=V, Cr, Fe, Co) are also $ i$-wave altermagnetic materials. Different from $\rm NiF_3$, the SOC effect of $\rm AF_3$ is weak near the Fermi level, which are demonstrated by our calculated band structures (Fig.\ref{fig:6}). The reason of $\rm AF_3$ with weak SOC effect is that valence electrons fully occupy all bands below the Fermi level (Table. \ref{tab:table1}). Thus, these altermagnetic materials $\rm AF_3$ are already semiconductors without SOC (Fig.\ref{fig:6}).

\begin{figure*}[htbp]
\centering
\includegraphics[width=0.90\textwidth]{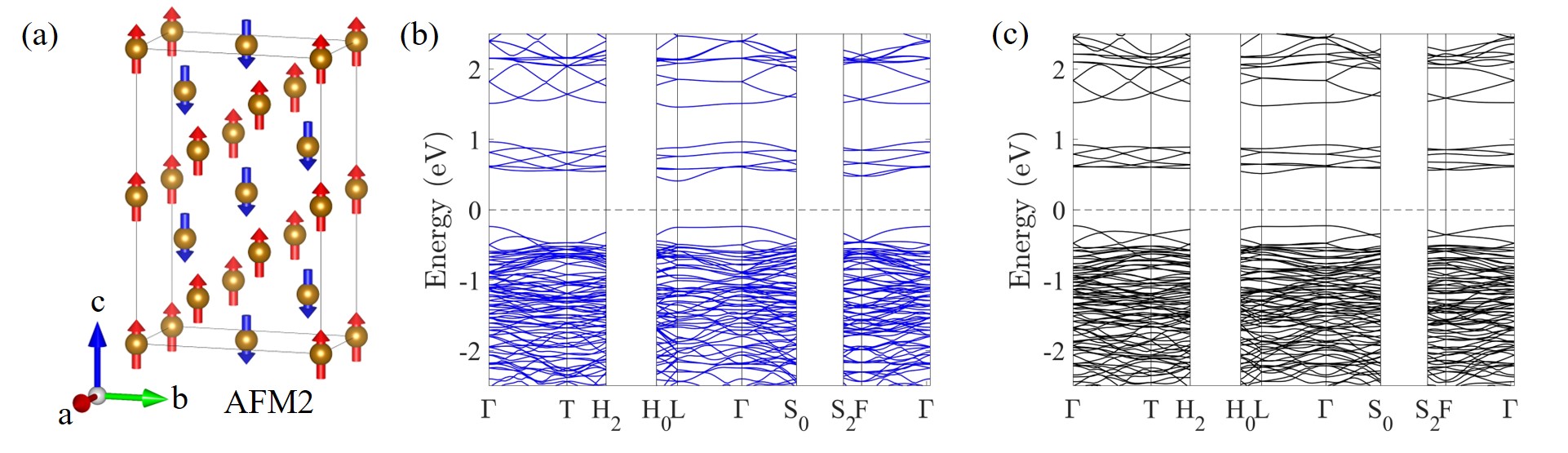}
\caption{(a) The magnetic structure of AFM2 for $\rm NiF_3$.  The band structure along the high-symmetry direction of AFM2 $\rm NiF_3$ without (b) and with (c) SOC.
}\label{fig:9}
\end{figure*}

On the other hand, the strong SOC in $\rm NiF_3$ closely related to altermagnetism. To demonstrate it, we have also calculated the band structure of $\rm NiF_3$ in the AFM2 antiferromagnetic state. Comparing the bands with and without SOC, AFM2 antiferromagnetic $\rm NiF_3$ does not have strong SOC effect (Fig.\ref{fig:9} (b) and (c)). 

The crystal structure and magnetic structure of $\rm FeCO_3$ are the same as $\rm NiF_3$ (Fig.\ref{fig:10}(a) and (b)), so $\rm FeCO_3$ is also an $ i$-wave altermagnetic material (Fig.\ref{fig:10}(c) and (e)). Interestingly, the quadruple degenerate band crossing the Fermi level is also half-filled for $\rm FeCO_3$ (Fig. \ref{fig:10}(d)). Moreover, our calculations show that the $\rm FeCO_3$ also has a strong SOC effect and its SOC effect is stronger than that of $\rm NiF_3$ (Fig.\ref{fig:10}(f)). Furthermore, the orbital weight analysis shows that the quadruple degenerate band crossing the Fermi level is all contributed by 3$ d$ orbitals of Fe atoms (Fig.\ref{fig:10}(d)), which is different from $\rm NiF_3$. Then, the enhanced SOC effect in $\rm FeCO_3$ is deriving from symmetry and the correlation interaction. On the other hand, the magnetic structure of $\rm FeCO_3$ has been confirmed by neutron scattering experiment \cite{r49}, so $\rm FeCO_3$ must have a strong SOC effect.   

\begin{figure*}[htbp]
\centering
\includegraphics[width=0.90\textwidth]{figure/fs7.jpg}
\caption{The crystal structure, magnetic structure and electronic band structures of $\rm FeCO_3$. The crystal structure (a) and magnetic structure (b) where the primitive cell of $\rm FeCO_3$ is given. (c) The anisotropic polarization charge densities. The red and blue represent spin-up and spin-down polarization charge density, respectively. Both (d) and (e) are the electronic band structures without SOC along the high-symmetry directions. The red and blue lines represent spin-up and spin-down bands, respectively. (f) The electronic band structure with SOC along the high-symmetry directions. The bandgap of $\rm FeCO_3$ is $\rm 3.49 eV$ under correlation interaction $\rm U = 5  eV$.
}\label{fig:10}
\end{figure*}

\subsection{Competition between the Jahn-Teller effect and SOC effect}

\begin{table*}[htbp]
\centering
\caption{\label{tab:table2}
 The relative energies of the high and low symmetric structures of $\rm NiF_3$}
\begin{tabular}{|c|c|r|}
\hline
$\rm U = 6.7 eV$ & without SOC (meV/Ni) & with SOC (meV/Ni) \\ \hline
High-symmetry crystal & $0.00$\,(metal) & $0.00$\,(insulator) \\\hline
Low-symmetry crystal  & $-388.82$\,(insulator) & $20.70$\,(insulator)\\\hline
\end{tabular}
\end{table*}

\begin{figure*}[htbp]
\centering
\includegraphics[width=0.90\textwidth]{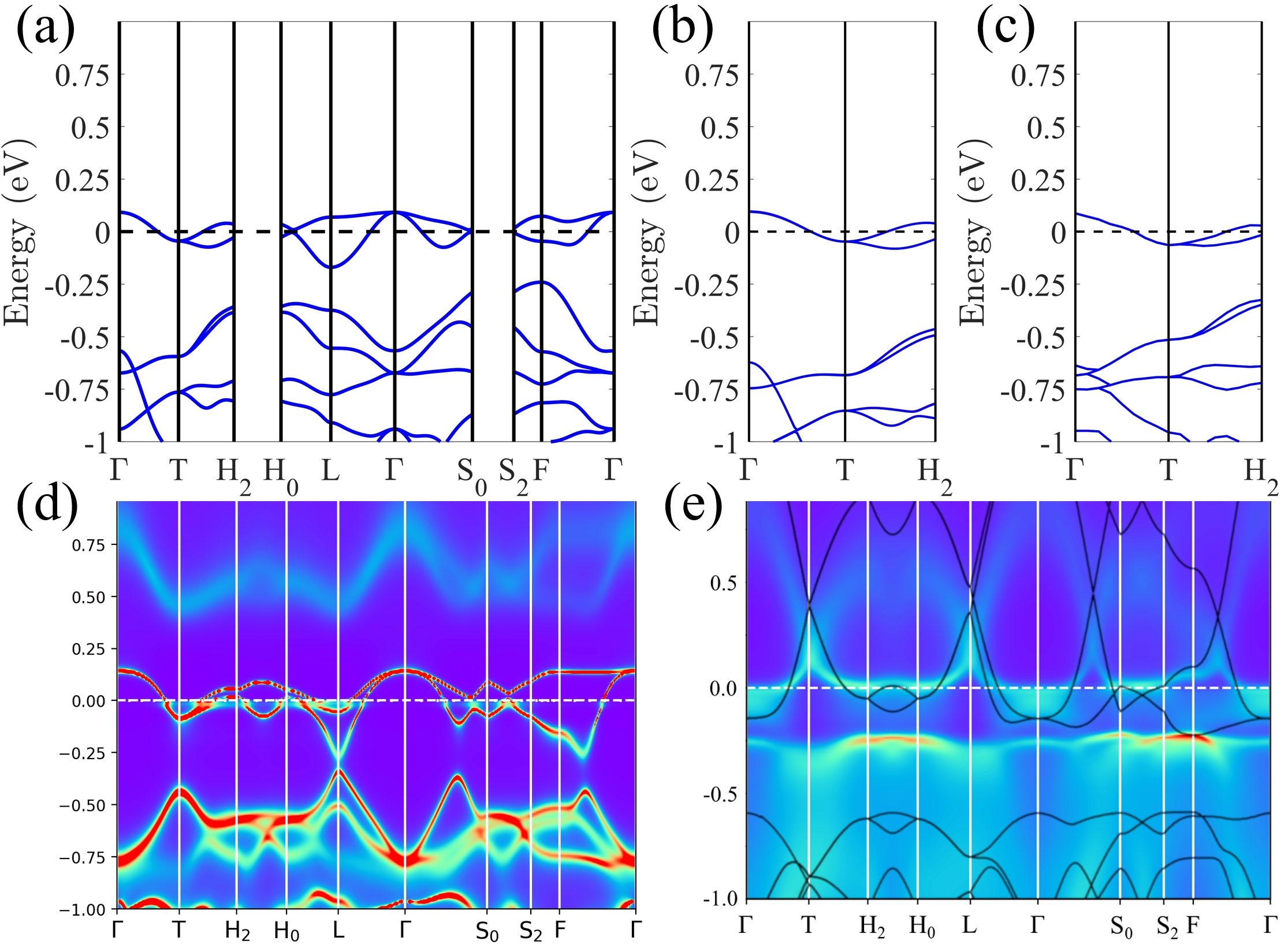}
\caption{The electronic band structures of altermagnetic $\rm NiF_3$ without SOC by GGA+U (a), SCAN (b), HSE06 (c) and DFT+DMFT (d) methods.
Momentum-resolved spectral function (e) obtained by the DFT+DMFT calculations and the black lines denote GGA+U band structure for nonmagnetic $\rm NiF_3$ without SOC.
}\label{fig:11}
\end{figure*}

The metal state of $\rm NiF_3$ is unstable without SOC. The stable insulating state of $\rm NiF_3$ can be achieved through the Jahn-Teller effect or SOC effect. In order to determine which interaction dominates, we first calculate the energies of high-symmetry and low-symmetry structures without SOC, and found that low-symmetry structure is more stable than the high-symmetry structure (Table. \ref{tab:table2}). Moreover, the $\rm NiF_3$ is an insulator under the low-symmetry structure. Interestingly, when SOC is considered, the stable phase changes from the low-symmetry structure to the high-symmetry structure (Table. \ref{tab:table2}). Furthermore, the low-symmetry structure can be relaxed to the high-symmetry structure under SOC. The SOC effect dominates the electronic properties of $\rm NiF_3$ rather than the Jahn-Teller effect. (The low-symmetry structure of $\rm NiF_3$ is obtained by shifting the position of the F atoms.) On the other hand, in order to make our results more reliable, we calculated the electronic structures of $\rm NiF_3$ without SOC by using DFT+U considered the on-site Coulomb repulsion U and Hund's coupling J, SCAN, HSE and GGA+DMFT methods, as shown in Fig.\ref{fig:11}. From Fig.\ref{fig:11}, $\rm NiF_3$ is always metal for all methods. In addition, we also do these calculations without SOC that we set two Ni magnetic moments with different sizes and opposite directions by using GGA+U methods, which are completely consistent with those calculations in the text (Fig.\ref{fig:2}). Therefore, the conclusion that there is an extremely strong spin-orbit coupling effect in the altermagnetic $\rm NiF_3$ is very reliable.

\subsection{Symmetry analysis}
The quadruple degenerate band of altermagnetic $\rm NiF_3$ on the $\rm \Gamma-T$ axis crossing the Fermi level is protected by spin group symmetry. According to the symmetry analysis in the text, the equivalent time-reversal spin symmetry $\rm \{C_2^\perp T||T\}$ results in two one-dimensional irreducible complex representations of the $\rm C_3$ symmetry to form a Kramers degeneracy, and the spin symmetry $\rm \{C_2^\perp ||M_1 t\}$ protects spin degeneracy. Thus, to achieve the quadruple degenerate band on the high-symmetry axis, the crystal symmetry needs to meet two conditions: first, the high-symmetry axis has $\rm C_3$, $\rm C_4$, $\rm C_6$ symmetries, and the $\rm C_3$, $\rm C_4$, $\rm C_6$ symmetries can’t have fractional translation; Second, the perpendicular to the high symmetry axis $\rm C_2$ symmetry or the mirror symmetry $\rm M$ containing the high symmetry axis has fractional translation, which can connect sublattices with antiparallel spins in altermagnetic materials. By symmetry analysis, we find that there are 34 nonsymmorphic space groups satisfying these two conditions, which are shown in Table. \ref{tab:table3}. In addition, sextuple degenerate band can be also realized in altermagnetic materials with nonsymmorphic space group symmetry, for example nonsymmorphic space group $\rm Pm\overline{3}n$.


\begin{table*}[htbp]
\caption{\label{tab:table3} Altermagnetic materials with quadruple degenerate bands along the high-symmetry direction need to satisfy the crystal space group symmetry. 
}
\begin{tabular}{|c|r|c|c|}
\hline
crystallographic system & Space group & Point group & Symmetry axis \\ \hline
Tetragonal &P4bm~(100) &$\rm C_{4v}$ &$\rm C_4$ \\\hline
Tetragonal &P4cc~(103) &$\rm C_{4v}$ &$\rm C_4$ \\\hline
Tetragonal &P4nc~(104) &$\rm C_{4v}$ &$\rm C_4$ \\\hline
Tetragonal &I4cm~(108) &$\rm C_{4v}$ &$\rm C_4$ \\\hline
Tetragonal &P4/mcc~(124) &$\rm D_{4h}$ &$\rm C_4$ \\\hline
Tetragonal &P4/mbm~(127) &$\rm D_{4h}$ &$\rm C_4$ \\\hline
Tetragonal &P4/mnc~(128) &$\rm D_{4h}$ &$\rm C_4$ \\\hline
Tetragonal &I4/mcm~(140) &$\rm D_{4h}$ &$\rm C_4$ \\\hline
Trigonal &P3c1~(158) &$\rm C_{3v}$ &$\rm C_3$ \\\hline
Trigonal &P31c~(159) &$\rm C_{3v}$ &$\rm C_3$ \\\hline
Trigonal &R3c~(161)  &$\rm C_{3v}$ &$\rm C_3$ \\\hline
Trigonal &$\rm P\overline{3}1c~(163)$ &$\rm D_{3d}$ &$\rm C_3$ \\\hline
Trigonal &$\rm P\overline{3}c1~(165)$ &$\rm D_{3d}$ &$\rm C_3$ \\\hline
Trigonal &$\rm R\overline{3}c~(167)$  &$\rm D_{3d}$ &$\rm C_3$ \\\hline
Hexagonal &$\rm P6_322~(182)$ &$\rm D_6$ &$\rm C_6$ \\\hline
Hexagonal &P6cc~(184) &$\rm C_{6v}$ &$\rm C_6$ \\\hline
Hexagonal &$\rm P6_3cm~(185)$ &$\rm C_{6v}$ &$\rm C_6$ \\\hline
Hexagonal &$\rm P6_3mc~(186)$ &$\rm C_{6v}$ &$\rm C_6$ \\\hline
Hexagonal &P6/mcc~(192) &$\rm D_{6h}$ &$\rm C_6$ \\\hline
Hexagonal &$\rm P6_3/mcm~(193)$ &$\rm D_{6h}$ &$\rm C_6$ \\\hline
Hexagonal &$\rm P6_3/mmc~(194)$ &$\rm D_{6h}$ &$\rm C_6$ \\\hline
Cubic &$\rm P4_232~(208)$ &$\rm O$ &$\rm C_3$ \\\hline
Cubic &$\rm F4_132~(210)$ &$\rm O$ &$\rm C_3$ \\\hline
Cubic &$\rm P4_332~(212)$ &$\rm O$ &$\rm C_3$ \\\hline
Cubic &$\rm P4_132~(213)$ &$\rm O$ &$\rm C_3$ \\\hline
Cubic &$\rm I4_132~(214)$ &$\rm O$ &$\rm C_3$ \\\hline
Cubic &$\rm P\overline{4}3n~(218)$ &$\rm T_d$ &$\rm C_3$ \\\hline
Cubic &$\rm F\overline{4}3c~(219)$ &$\rm T_d$ &$\rm C_3$ \\\hline
Cubic &$\rm I\overline{4}3d~(220)$ &$\rm T_d$ &$\rm C_3$ \\\hline
Cubic &$\rm Pm\overline{3}n~(223)$ &$\rm O_h$ &$\rm C_3$ \\\hline
Cubic &$\rm Fm\overline{3}c~(226)$ &$\rm O_h$ &$\rm C_3$ \\\hline
Cubic &$\rm Fd\overline{3}m~(227)$ &$\rm O_h$ &$\rm C_3$ \\\hline
Cubic &$\rm Fd\overline{3}c~(228)$ &$\rm O_h$ &$\rm C_3$ \\\hline
Cubic &$\rm Ia\overline{3}d~(230)$ &$\rm O_h$ &$\rm C_3$ \\\hline
\end{tabular}
\end{table*}




\end{appendices}


\end{document}